\documentclass[12pt]{article}
\pdfoutput=1

\usepackage[utf8]{inputenc}

\usepackage{draft}
\usepackage{putex}

 \usepackage{dsfont}
\usepackage[all]{xy}

\usepackage{stackrel,amssymb}

\usepackage{graphicx}
\usepackage{caption}
\usepackage{amsmath,bm}
\usepackage{url}
\usepackage{array}
\usepackage{subcaption}
\usepackage{epstopdf}
\usepackage{enumerate}
\usepackage{cite}
\usepackage{tensor}
\usepackage{slashed}
\usepackage[utf8]{inputenc}
\usepackage{rotating}
\usepackage{bigfoot}
\usepackage[
      colorlinks=true,
      linkcolor=blue,
      urlcolor=blue,
      filecolor=black,
      citecolor=red,
      ]{hyperref}

\usepackage{tikz-cd}

\def\XXint#1#2#3{{\setbox0=\hbox{$#1{#2#3}{\int}$ }
		\vcenter{\hbox{$#2#3$ }}\kern-.6\wd0}}

  \usepackage{adjustbox}
\usepackage{multirow}
\usepackage{tikz-cd}

\numberwithin{equation}{section}

\newtheorem{theorem}{Theorem}
\newtheorem{conj}{Conjecture}

\def\<{\langle}
\def\>{\rangle}

\def\pa{\partial}
\def\ve{\varepsilon}

\newcommand{\leftrarrows}{\mathrel{\raise.75ex\hbox{\oalign{%
				$\scriptstyle\leftarrow$\cr
				\vrule width0pt height.5ex$\hfil\scriptstyle\relbar$\cr}}}}
\newcommand{\lrightarrows}{\mathrel{\raise.75ex\hbox{\oalign{%
				$\scriptstyle\relbar$\hfil\cr
				$\scriptstyle\vrule width0pt height.5ex\smash\rightarrow$\cr}}}}
\newcommand{\Rrelbar}{\mathrel{\raise.75ex\hbox{\oalign{%
				$\scriptstyle\relbar$\cr
				\vrule width0pt height.5ex$\scriptstyle\relbar$}}}}

\makeatletter
\def\leftrightarrowsfill@{\arrowfill@\leftrarrows\Rrelbar\lrightarrows}
\newcommand{\xleftrightarrows}[2][]{\ext@arrow 3399\leftrightarrowsfill@{#1}{#2}}
\makeatother

\begin{document}

\preprint{CALT-TH 2024-015, IPMU 24-0011}

\title{Universal Bounds on CFT Distance Conjecture}

\authors{Hirosi Ooguri\worksat{\CALTECH, \IPMU} and Yifan Wang\worksat{\NYU}}

\institution{CALTECH}{Walter Burke Institute for Theoretical Physics, California Institute of Technology, Pasadena, CA
91125, USA}

\institution{IPMU}{Kavli Institute for the Physics and Mathematics of the Universe (WPI), University of Tokyo,
Kashiwa 277-8583, Japan}

\institution{NYU}{Center for Cosmology and Particle Physics, New York University, New York, NY 10003, USA}

\abstract{
For any unitary conformal field theory in two dimensions with the central charge $c$,  we prove that, if there is a nontrivial primary operator whose conformal dimension $\Delta$ vanishes in some limit on the conformal manifold, the Zamolodchikov distance $t$ to the limit is infinite, the approach to this limit is exponential $\Delta = \exp(- \alpha t +O(1) )$, and the decay rate obeys the universal bounds $c^{-1/2} \leq \alpha \leq 1$. In the limit, we also find that an infinite tower of primary operators emerges without a gap above the vacuum and that the conformal field theory becomes locally a tensor product of a sigma-model in the large radius limit and a compact theory.
As a corollary, we establish a part of the Distance Conjecture about gravitational theories in three-dimensional anti-de Sitter space. In particular, our bounds on $\alpha$ indicate that the emergence of exponentially light states is inevitable as the moduli field corresponding to $t$ rolls beyond the Planck scale along the steepest path and that this  phenomenon  can begin already at the curvature scale of the bulk geometry. We also comment on implications of our bounds for gravity in asymptotically flat spacetime by taking the flat space limit and compare with the Sharpened Distance Conjecture.
}
\date{}

\maketitle

\tableofcontents

\section{Introduction and Summary}
Over the past couple of decades, it has become increasingly clear that there are constraints on the low-energy effective theories of quantum gravity that cannot be captured by the standard Wilsonian paradigm. These constraints delineate the boundary between the Landscape and the Swampland \cite{Vafa:2005ui}. For gravitational theories in asymptotically anti-de Sitter (AdS) spacetimes, we can formulate such constraints and aim to prove or falsify them using the AdS/CFT correspondence. For example, it was proven in \cite{Harlow:2018jwu,Harlow:2018tng} that any global symmetry in a quantum gravity theory in AdS would lead to an inconsistency in its dual conformal field theory (CFT). 

The Distance Conjecture \cite{Ooguri:2006in} has been one of the most well-tested among Swampland conditions. The conjecture claims the following set of properties about continuous moduli of quantum gravity theories, starting with:

\setcounter{conj}{-1}
\begin{conj} 
 The moduli space ${\cal M}$ is parametrized by expectation values of massless scalar fields.
 \label{conj:zero}
\end{conj}
 If this conjecture holds, the moduli space is endowed with a natural metric given by the kinetic term of the moduli fields, which defines a notion of distance $d(p,p')$ between any two points $p,p'\in \cM$. Among other conjectures formulated in  \cite{Ooguri:2006in} are:

\begin{conj}
    Choose any point $p_0 \in {\cal M}$. For any positive $t$, there is another point $p \in {\cal M}$  such that the distance $d(p, p_0)$ between $p$ and $p_0$ is greater than $t$.
\label{conj:one}
 \end{conj} 

\begin{conj}
Compared to the theory at $p_0 \in {\cal M}$, for sufficiently large $t$, the theory at $p$ with $d(p, p_0) > t$ has an infinite tower of light particles starting with mass of the order of $\exp(- \alpha t)$ for some $\alpha > 0$. In the $t \rightarrow \infty$ limit, the number of extra light particles of mass less than a fixed mass scale becomes infinite.
 \label{conj:two}
\end{conj}

For gravitational theories in AdS, Conjecture \ref{conj:zero} can be shown as follows. If there is a continuous parameter in AdS, there is a corresponding parameter in the dual CFT. Such a parameter is believed to be associated with an exactly marginal operator in the CFT, which then corresponds to either a massless scalar field in the bulk or (when the marginal operator is double-trace) a continuous deformation of the boundary condition at the infinity of AdS. In particular, continuous parameters in the bulk Lagrangian must be expectation values of massless scalar fields. 

However, this conjecture alone does not lead to a sharp constraint on a low-energy effective Lagrangian, since the parameters of the Lagrangian may have been fixed at high energy, $e.g.$, by potentials for the corresponding scalar fields. This is analogous to the absence of global symmetry \cite{Harlow:2018jwu,Harlow:2018tng},
which also does not produce a sharp constraint on a low-energy effective Lagrangian since the low-energy theory may have an accidental symmetry, which is broken or gauged at high energy. The analogy can be made more precise by interpreting Conjecture \ref{conj:zero} as the absence of $(-1)$-form global symmetry. On the other hand, if $\alpha$ can be bounded, Conjecture \ref{conj:two} will give a sharp constraint on low-energy effective theories.

The AdS versions of Conjecture \ref{conj:one} and \ref{conj:two} have been proposed in \cite{Perlmutter:2020buo} for bulk spacetime dimensions  $\geq 4$. The main claim is that all theories at infinite distance in the bulk moduli space have an emergent higher spin symmetry, generated by an infinite tower of conserved currents. Since the bulk moduli space is identified as a conformal submanifold of the dual CFT, which we will also denote as $\cM$, these conjectures can be stated precisely in CFT terms and therefore are called the CFT Distance Conjectures.
They include: 

\setcounter{conj}{0}
\renewcommand{\theconj}{\Roman{conj}}
\begin{conj}
    All points with emergent higher spin symmetries are at infinite distance on $\cM$.
    \label{conj:CFTI}
\end{conj}

\begin{conj} All CFTs at infinite distance on $\cM$ are higher spin points.
\label{conj:CFTII}
\end{conj}

For supersymmetric theories, Conjecture~\ref{conj:CFTI} was proven in \cite{Perlmutter:2020buo} by using the fact that higher spin symmetries imply the existence of free decoupled sectors in their dual superconformal theories. More recently, it was proven for any unitary CFT with an energy-momentum tensor in \cite{Baume:2023msm}. Conjecture~\ref{conj:CFTII} remains open. 

For CFT in two dimensions, these conjectures need to be modified since there are always higher spin currents constructed from composites of the holomorphic stress tensor \cite{Kontsevich:2000yf,Roggenkamp:2003qp,MR2742431,Perlmutter:2020buo}. In this paper, we prove the following four theorems about two-dimensional CFTs.

\begin{theorem} 
If there is a geodesic on the conformal manifold ${\cal M}$ along which the conformal dimension $\Delta$ of a primary operator
vanishes in some limit, then
the geodesic distance $t$ to the limit measured by the Zamolodchikov metric is infinite.
\label{thm:zerogaptoinfdist}
\end{theorem}

\begin{theorem}  In the limit, $\Delta$ 
vanishes exponentially as $\Delta = \exp(- \alpha t + O(1))$
with the universal upper bound $\alpha \leq 1$.
\label{thm:gapvanishing}
\end{theorem}

\begin{theorem}
The compact CFT of central charge $c$ in the limit of vanishing $\Delta$ contains a subalgebra of local operators which are described by the sigma-model on $\mR^N$ for some positive integer $N \leq c$.
\label{thm:CFTinthelimit}
\end{theorem}
Theorem \ref{thm:CFTinthelimit} shows that the limit can always be understood as the decompactification limit of an emergent target space of CFT and confirms the conjecture of Kontsevich and Soibelman in \cite{Kontsevich:2000yf}. 

In general, the parameter $\alpha$ defined in Theorem~\ref{thm:gapvanishing} depends on the geodesic to reach the limit as well as on the primary operator we follow along the geodesic. For the optimal choice of geodesic (which we assume to be in the direction of a parity-even exactly marginal operator) and primary operator, we can derive the following lower bound on $\alpha$,
\begin{theorem}
There exists a geodesic and a primary operator with a vanishing conformal dimension along the geodesic such that the exponential decay rate obeys $N^{-1/2} \leq \alpha$.\label{thm:lowerbound}
\end{theorem}
Since $N\leq c$, this theorem also implies the lower bound $c^{-1/2}\leq \alpha$.Combining these results, we obtain the upper and lower bounds on $\alpha$,   
\begin{equation}
    \frac{1}{\sqrt{c}} \leq \alpha \leq 1 \, .
    \label{theinequality}
\end{equation}
These bounds are sharp, and we will find necessary and sufficient conditions to saturate each bound at $\alpha=c^{-1/2}$ and $\alpha=1$. 
 For superconformal CFTs, the lower bound is strengthened to $\sqrt{3/2} \ c ^{-1/2}\leq \alpha$.

To prove these four theorems, we do not need to assume that the CFTs have holographic duals in AdS or that the central charge $c$ is large. We only assume that the CFTs are unitary and each have a normalizable conformally invariant vacuum (away from the limit), there is an exactly marginal operator for each tangent vector on their conformal manifolds, and the genus-zero four-point functions of the light primary operators are well-defined in the limit of vanishing gap $\Delta_{\rm gap} \rightarrow 0$.
To prove Theorems~\ref{thm:zerogaptoinfdist} and \ref{thm:gapvanishing}, 
we do not even assume the existence of a local stress tensor. Therefore, these theorems also apply to the conformal manifolds of surface defects in $d\geq 3$ CFTs (such as the Gukov-Witten surface defects in the $\cN=4$ super-Yang-Mills theory \cite{Gukov:2006jk}). 

Furthermore, Theorem~\ref{thm:CFTinthelimit} does not assume that the family of CFTs is related by deformation with an exactly marginal operator. Therefore, it also holds for a discrete sequence of CFTs under the assumptions stated in the above. 
For example, the large $k$ limit of the level $k$ Wess-Zumino-Witten model for a compact Lie group $G$ is locally equivalent to the theory of free non-compact bosons  with $c = \dim G$, and the large $k$ limit of the $A_k$-type Virasoro minimal model CFTs is locally described in terms of a non-compact boson at $c=1$
with a pair of walls in the target space infinitely distant from each other \cite{Runkel:2001ng, Mazel:2024alu}.  If we can also 
generalize Theorems~\ref{thm:zerogaptoinfdist},~\ref{thm:gapvanishing}, and~\ref{thm:lowerbound} to include discrete families of CFTs,  it would open the possibility to study the flat space limits of AdS gravities and test the conjecture in \cite{Lust:2019zwm} (see also related discussions in Section~\ref{sec:gravityimplications}).

\subsection{General Structure of Conformal Manifolds}
Let us review general properties of conformal manifolds for $d$-dimensional CFTs relevant for this paper. We take $\lambda^i$ to be the coordinates on the conformal manifold $\cM$. Locally on $\cM$, the CFT action reads
\ie 
S(\lambda+\D \lambda)=S(\lambda)+ {1\over V_{d-1}} \int d^d x \, \D \lambda^i M_i(x)\,,
\label{Sdef}
\fe
where $M_i$ are exactly marginal operators labeling the tangent directions on $\cM$ and $V_{d-1}$ is the volume of the unit sphere $S^{d-1}$.

It is understood that
the conformal manifold $\cM$ is endowed with a natural Riemannian metric, namely the Zamolodchikov metric \cite{Zamolodchikov:1986gt}, defined by the two-point functions of $\cO_i$,
\ie 
g_{ij}(\lambda)=|x|^{2d}\la M_i(x) M_j(0)\ra_{\lambda}\,,
\label{zmetric}
\fe
which is manifestly positive definite as a consequence of unitarity. Note that \eqref{Sdef} fixes a \textit{canonical} normalization for the distance on $\cM$ measured by \eqref{zmetric}.

The conformal manifold $\cM$ may have singularities due to orbifold quotients by duality groups \cite{dijkgraaf1988moduli,Seiberg:1988pf}, divergent Riemann curvature \cite{Candelas:1990rm} and emergent exactly marginal operators at special loci \cite{Dijkgraaf:1987vp,Ginsparg:1987eb,Dixon:1988qd}. Furthermore, the conformal manifold  $\cM$ is naturally geodesically complete with respect to \eqref{zmetric}, yet in general non-compact.
The non-compact directions where the geodesic distance diverges give rise to infinite distance limits of $\cM$.

In addition to these
intrinsic geometric features, the conformal manifold $\cM$ hosts a great deal of extra structure by consideration of how CFT data varies with respect to $\lambda^i$, which abstractly define certain fiber bundles over $\cM$. It is natural to ask if and how such structure are constrained by the intrinsic geometry on $\cM$ (see for example \cite{Cecotti:1991me,Bershadsky:1993ta,deBoer:2008ss,Baggio:2014ioa,Gomis:2015yaa,Gomis:2016sab,Seiberg:2018ntt,Balthazar:2022hzb}). Here we will focus 
on the interplay with infinite distance limits on  $\cM$.

Of particular interest are universal quantities such as the conformal dimension gap in the CFT $\cT_\lambda$ at a point $\lambda$ on the conformal manifold $\cM$,
\ie 
\Delta_{\rm gap}(\lambda)\equiv \inf \{ \Delta \,|\,(\Delta,j_a)\in \cT_\lambda \}  \,,
\label{Dgap}
\fe 
where $\Delta$ labels the scaling dimension and $j_a$ labels the representation of SO$(d)$ in terms of spins on the $\left\lfloor d/ 2 \right\rfloor$ orthogonal two-planes for the local operators at $\cT_\lambda$.
Similarly for operators with sufficiently high spin, we define the twist gap $t_{\rm gap}$,
\ie 
t_{\rm gap}(\lambda;j_a) =  \inf \{  \Delta -\sum_{a=1}^{\left\lfloor d/ 2 \right\rfloor} |j_a|   \,|\,(\Delta,j_a)\in \cT_\lambda \} \}\,.
\label{tgap}
\fe
These quantities offer a glimpse of how the spectrum of local operators vary over $\cM$. Note that both quantities are non-negative and bounded from below by the corresponding unitarity bounds. 
It has been conjectured in \cite{Kontsevich:2000yf,MR2742431,Perlmutter:2020buo} that the behavior of these quantities approaching their unitarity bounds are correlated with infinite distance limits on $\cM$.\footnote{More precisely, in \cite{Kontsevich:2000yf,MR2742431}, the metric on the conformal manifold is not specified, rather an analogy was made to the Gromov-Hausdorff metric for a family of Riemannian manifolds based on the connection with $d=2$ CFTs defined via sigma-models. In a similar way, the degenerating $d=2$ CFTs were compared to the degeneration limits of the manifolds. In particular, it is conjectured in \cite{Kontsevich:2000yf,MR2742431} that the subspace of $\cM$ where the dimension gap is bounded from below $\Delta_{\rm gap}(\lambda)\geq \ve$ for $\ve>0$ is compact in a suitably defined Gromov-Hausdorff (metric) topology. Furthermore, by including the degenerating limits, one can define a compactification $\overline \cM$ of $\cM$ in the Gromov-Hausdorff topology. See \cite{gross2000large,zhang2016completion} for explicit examples.} For CFTs in dimension $d\geq 3$, it was recently proven in \cite{Baume:2023msm} that the vanishing of $t_{\rm gap}(\lambda;j_a)-(d-2)$ for operators in the symmetric traceless representations of SO$(d)$ of high ranks (also known as higher-spin operators) directly implies infinite distance on $\cM$. While the converse statement has not been proven, criterion on infinite versus finite distance on $\cM$ was provided in terms of the CFT data in \cite{Baume:2023msm}.
 
In this work, we study conformal manifolds of CFTs in dimension $d=2$ and investigate  infinite distance limits on the conformal manifold $\cM$ and corresponding behavior of the CFT data. 
As was already noted in \cite{Kontsevich:2000yf,Roggenkamp:2003qp,MR2742431,Perlmutter:2020buo}, because of the enhancement of the conformal symmetry to Virasoro symmetry in $d=2$,
instead of the twist gap \eqref{tgap}, the dimension gap \eqref{Dgap} is potentially the universal indicator for infinite distance on $\cM$. Indeed for bosonic CFTs defined by a sigma-model on the flat torus $T^n=\mR^n/\Lambda$ for integral lattice $\Lambda$, this correspondence between vanishing $\Delta_{\rm gap}$ and infinite distance on the Narain moduli space is immediate: in a fixed T-dual frame,  infinite tower of momentum operators develop vanishing dimensions along directions of $T^n$ that decompactify. 
More generally, for CFTs defined by $\cN=(2,2)$ supersymmetric sigma-models on compact Calabi-Yau manifolds, infinite operators come down to vanishing dimensions in a large volume limit of the target manifold (equivalently in a large complex structure limit of the mirror target manifold). Furthermore it is conjectured in \cite{Kontsevich:2000yf} that such degeneration limits of CFT (where $\Delta_{\rm gap}$ vanishes) are always associated with emergent geometries that describe a closed sector of the full CFT.
A main goal of the paper is to argue that these are indeed universal phenomena in $d=2$ CFTs, regardless of whether a sigma-model description exists in the interior of the conformal manifold. The main results are summarized in Theorem~\ref{thm:zerogaptoinfdist},~\ref{thm:gapvanishing},~\ref{thm:CFTinthelimit}, and~\ref{thm:lowerbound}.
  
We emphasize that the CFT in the limit $\Delta_{\rm gap}\to 0$ does not obey the usual axioms of a compact unitary CFT ($e.g.,$ the genus-one partition function diverges) but we assume the 
sphere correlation functions of light operators are well-defined in this limit. 
Degeneration limits of this type have been discussed in \cite{Kontsevich:2000yf} and later defined in precise CFT language in \cite{Roggenkamp:2003qp} but it is not established if such a limit exists in general. Our findings further clarify features of the limiting theory and provide a consistency check on the assumption of well-defined sphere correlators in this limit.

\subsection{Constraints on Gravitational Theories in AdS$_3$ and Flat Space}
\label{sec:gravityimplications}

For the purpose of understanding the implications of our CFT results for quantum gravity, 
it is useful to 
translate  the bounds \eqref{theinequality} on $\alpha$ 
in the units appropriate for a gravitational theory in AdS$_3$. If we normalize the kinetic term of the massless scalar field $\phi$ in AdS$_3$ dual to the geodesic coordinate $t$ on the conformal manifold of CFT$_2$ as
\begin{equation}
    {\cal L} = \frac{1}{2} (\partial \phi)^2 + \cdots\,,
    \label{canscalar}
\end{equation}
without the inverse of the Newton constant in front, by the AdS/CFT dictionary 
we can identify the asymptotic value of $\phi$ with the geodesic distance  $t$  on the conformal manifold as $\phi =t \cdot (8\pi L_{\rm AdS})^{-1/2}$, where $L_{\rm AdS}$ is the curvature radius of AdS  \cite{Freedman:1998tz,DHoker:2002nbb}.\footnote{This follows by comparing the normalization of the exactly marginal operator that couples to $\phi$ on the AdS boundary and that to the proper distance $t$ defined with respect to the Zamolodchikov metric \eqref{zmetric},
\ie 
{(2\Delta-d)\Gamma(\Delta)\over \pi^{d\over 2}\Gamma(\Delta-d/2)}{ L_{\rm AdS}^{d-1}\phi^2 } = {t^2\over V_{d-1}^2} \,,
\fe
with $\Delta=d$ where the LHS follows from standard bulk computation \cite{DHoker:2002nbb} and the RHS is a consequence of our normalization of the deformation \eqref{Sdef}.
} Correspondingly, $\alpha_{\rm AdS} = \alpha\cdot  (8\pi L_{\rm AdS})^{1/2}$ controls the exponential decay $e^{-\A_{\rm AdS} \phi}$  of the \textit{energy} gap in the bulk.
Using the relation
\begin{equation}
c= \frac{3 L_{\rm AdS}}{2G_N}= 8 \pi L_{\rm AdS} \cdot \frac{3}{2 L_{\rm Planck}}
\,,
\end{equation}
where  $G_N$ is the Newton constant and $L_{\rm Planck}=8 \pi G_N$ is the reduced Planck length 
in three dimensions.
The inequality (\ref{theinequality}) can then be expressed in terms of the bulk variables as, 
\begin{equation}
\left(\frac{2}{3}L_{\rm Planck}\right)^{1/2}
 \leq \ \alpha_{\rm AdS} \ \leq 
\left( 8\pi L_{\rm AdS}\right)^{1/2} \,.
\label{AdSgravbound}
\end{equation}
The lower bound means that the emergence of  exponentially light states with energy $\Delta$ is inevitable 
when $\phi$ rolls beyond the Planck scale at $\phi = 
(2 L_{\rm Planck}/3)^{-1/2}$ along the path of the steepest descent, while the upper bound implies that this phenomenon can  begin already at the AdS curvature scale $\phi = 
(8\pi L_{\rm AdS})^{-1/2}$. 

If the theory is supersymmetric, the lower bound is strengthened as stated below \eqref{theinequality}, and can be translated in the AdS units as 
\begin{equation}
\left(L_{\rm Planck}\right)^{1/2}
 \leq \ \alpha_{\rm AdS} \ \leq 
\left( 8\pi L_{\rm AdS}\right)^{1/2} \,.
\label{AdSgravboundSUSY}
\end{equation}

These bounds on the decay rate of the AdS energy can be translated to that of the mass for the corresponding particles using the standard AdS/CFT dictionary \cite{DHoker:2002nbb}.  
The energy $\Delta$ and the mass $m$ for scalar fields in AdS$_3$ are related by 
\ie 
\Delta = 1 \pm \sqrt{1+m^2L_{\rm AdS}^2}\,,
\fe
where the plus sign corresponds to the standard quantization and the minus sign corresponds to the alternative quantization, which is valid for $m^2<0$. Furthermore, unitarity requires  the Breitenlohner-Freedman (BF) bound, $m^2 L_{{\rm AdS}}^2\geq -1$
\cite{Breitenlohner:1982jf}. The light scalar particle states arise from the alternative quantization of the scalar field dual to the light scalar operators in the CFT.
Therefore, while the energy of these scalar particle states approach zero from above, their masses-squared approach zero from below as in 
\ie
m^2 \sim - \frac{2}{L_{{\rm AdS}}^2 }\Delta
= - \frac{2}{L_{{\rm AdS}}^2 }
\exp(-\A_{\rm AdS} \phi + O(1))\,,
\fe
which is a phenomena specific to gravity in AdS$_3$ at infinite distance on the moduli space (since unitarity bounds in higher dimensions are stronger). Note that,
in addition to the exactly marginal operator corresponds to $t$, 
other marginal operators may emerge at infinite distance (see Section~\ref{sec:limitingCFT}). They correspond to scalar fields in AdS$_3$ with the standard quantization and their masses-squared approaching zero from above.

Furthermore, there are also infinite towers of massive spinning particles whose masses-squared approach zero from above in the limit, in a way similar to what happens in higher dimensional AdS gravity as described in \cite{Perlmutter:2020buo}. Indeed they are associated with massive (higher) spin fields which correspond to (higher) spin (quasi-primary) operators in the CFT that only become conserved in the limit ($i.e.$ vanishing twist). In this case, the dictionary between the energy (scaling dimension) and the mass is 
\ie 
\Delta = 1+ \sqrt{(s-1)^2+ m^2 L_{{\rm AdS}}^2 }
\fe
for $s\geq 1$ (and the alternative quantization is only possible for $s=1$ at $m=0$). Therefore, for the massive particles with spin $s\geq 2$ whose energy approaches the unitarity bound $\Delta=s$ in the limit, their masses-squared approach zero from above as for the twist in 
\ie
m^2 \sim \frac{2(s-1)}{L_{{\rm AdS}}^2 }(
\Delta - s) = \frac{2(s-1)}{L_{{\rm AdS}}^2 }\exp(-\A_{\rm AdS} \phi + O(1))
\,.
\label{twistgap}
\fe
For the special case of massive vector ($i.e.$ $s=1$), this becomes
\ie
|m|= \frac{1}{L_{{\rm AdS}} }(
\Delta - 1) = \frac{1}{L_{{\rm AdS}}} \exp(-\A_{\rm AdS} \phi + O(1))
\,.
\label{twistgap1}
\fe
In the dual CFT, such higher spin particles correspond to spinning operators constructed from products of the emergent currents in the infinite distance limit\footnote{They are Virasoro primary operators (and their quasi-primary descendants) that are descendants with respect to the emergent current algebra in the limit. See Section~\ref{sec:limitingCFT} for details.} and this is how the last equalities in \eqref{twistgap} and \eqref{twistgap1} are deduced.

Let us now comment on the implications of our bounds in AdS on the decay rates of the masses of light particles for the similar question of quantum gravity in asymptotically flat space ($i.e.$ Conjecture~\ref{conj:two} in flat space). To this end, we note that a lower bound on the mass decay rate in the effective field theory in $D$-dimensional flat space was proposed in \cite{Etheredge:2022opl}\footnote{We thank José Calderón-Infante for bringing this paper to our attention.} under the name of the Sharpened Distance Conjecture. It claims the existence of an infinite tower of light particles with exponentially descreasing mass of the order $\exp(-\alpha_{{\rm flat}, D} \tilde \phi)$ with the coefficient $\alpha_{{\rm flat}, D}$ bounded below as,
 \ie 
\alpha_{{\rm flat}, D} \geq  {L_{{\rm Planck},D}^{(D-2)/2} \over (D-2)^{1/2} }\,,
\label{SDC}
\fe 
where  $L_{{\rm Planck},D}$ is the reduced Planck length in $D$ dimensions and
the modulus field $\tilde \phi$ is canonically normalized  as in \eqref{canscalar}.
As explained in \cite{vandeHeisteeg:2023ubh}, 
this bound follows from the 
Emergent String Conjecture \cite{Lee:2019wij}, which states that an infinite distance limit of its moduli space either decompactifies, or reduces to an asymptotically tensionless, weakly coupled string theory. This bound was further supported by evidence from supergravity \cite{Etheredge:2022opl}.

By taking an appropriate flat space limit $L_{\rm AdS}\to \infty$ (whose details such as the scaling of other parameters in the putative limit depend on the theory \cite{Polchinski:1999ry,Susskind:1998vk,Giddings:1999jq}), our bounds for the AdS$_3$ gravity should produce bounds for quantum gravity in asymptotically flat spacetime of dimension $D=3+n\geq 3$ where $n$ counts the internal dimensions that decompactify in this limit.\footnote{
It was conjectured in \cite{Lust:2019zwm} that such a decompactifying internal space always exists in any AdS gravity. Though the KK modes on the decompactifying internal space also become massless in the flat space limit, we are interested in the contribution to the mass that do not vanish in this limit.} 
Our lower bound \eqref{AdSgravbound}, after taking into account the canonical normalization in $D$ dimensions,\footnote{Since both $\tilde \phi$ and $\phi$ are canonically normalized in $D$ and $3$ dimensions, respectively, they are related as 
$\phi = V_{D-3}^{1/2} \ \tilde \phi$, where $V_{D-3}$ is the volume of the internal dimensions that decompactify in the limit. Therefore $\alpha_{\rm AdS} = V_{D-3}^{-1/2}\ \alpha_{{\rm flat}, D}$ in the flat space limit. On the other hand, $L_{{\rm Planck}, 3}= L_{{\rm Planck},D}^{D-2}/V_{D-3}$. Combining these, \eqref{AdSgravbound} gives \eqref{flatbound} in the flat space limit.} implies that
\ie 
\alpha_{{\rm flat}, D} \geq {1\over \sqrt{6}} L_{{\rm Planck},D}^{(D-2)/2}\,,
\label{flatbound}
\fe
and for the supersymmetric case a stronger lower bound follows from \eqref{AdSgravboundSUSY},
\ie 
\alpha_{{\rm flat}, D} \geq {1\over 2} L_{{\rm Planck},D}^{(D-2)/2}\,.
\label{flatlowerboundSUSY}
\fe
Naively, these bounds can be further strengthened by a factor of 2, due to the faster decay of the mass of spin $s=1$ particles in \eqref{twistgap1}. However, from known top-down examples, we expect these vector fields to be governed by a one-derivative action to leading order in AdS$_3$ and their mass-squared in the flat space limit obey the same exponential decay rate as the other light particles.\footnote{In particular, by continuity, the massless limit of these massive vector field is governed by a Chern-Simons action as derived in \cite{Hansen:2006wu}.} We emphasize that taking the flat space limit generally involves a discrete sequence of AdS/CFT dual pairs ($e.g.$ with increasing central charge $c$) \cite{Polchinski:1999ry,Susskind:1998vk,Giddings:1999jq}, and when deducing the bounds above we have assumed that the direction parametrized by $\phi$ (or $t$) on the  AdS moduli space (CFT conformal manifold) is common to all instances in the sequence. 
  
Curiously the lower bound \eqref{flatlowerboundSUSY} for supersymmetric theories agrees
with the Sharpened Distance Conjecture  \eqref{SDC} at $D=6$  \cite{Etheredge:2022opl}.  Indeed, well-known supersymmetric ${\rm AdS}_3/{\rm CFT}_2$ examples such as type IIB string theory on  ${\rm AdS}_3\times {\rm S}^3\times M$ for $M=T^4$ or $M=K3$ dual to the D1-D5 CFT \cite{Maldacena:1998bw,deBoer:1998kjm} are described by 6d supergravity in the flat space limit.\footnote{Other supersymmetric examples include type IIB string theory on ${\rm AdS}_3\times {\rm S}^3\times {\rm S}^3 \times {\rm S}^1$ \cite{Elitzur:1998mm,deBoer:1999gea,Gukov:2004ym,Tong:2014yna,Eberhardt:2017pty} which has a 9d flat space limit and M-theory on ${\rm AdS}_3\times {\rm S}^2\times {\rm CY}_3$ \cite{Maldacena:1997de,Minasian:1999qn,Denef:2007yt,deBoer:2008fk} for a compact Calabi-Yau three-fold denoted ${\rm CY}_3$ which has a 5d flat space limit.}    
On the other hand, the Sharpened Distance Bound 
may be violated without supersymmetry,
as we will discuss at the end of Section~\ref{sec:limitingCFT}. Below we will provide more details on this agreement between our result \eqref{flatlowerboundSUSY} and that of \cite{Etheredge:2022opl} in the supersymmetric setting using the aforementioned example of the D1-D5 CFT.

In this case, the CFT is constructed from a system of parallel $N_1$ D1 branes and $N_5$ D5 branes and described 
by a supersymmetric sigma-model on the symmetric orbifold ${\rm Sym}^{N_1 N_5}(M)$
with central charge (see $e.g.$ \cite{Maldacena:1998bw} for details)
\ie 
c=6N_1 N_5\,.
\fe
The length scales in the AdS dual are determined as follows,
\ie 
L_{\rm AdS}=(N_1 N_5)^{1\over 4} g_6^{1\over 2} \ell_s\,,\quad 
L_{{\rm Planck},3}=2\pi (N_1 N_5)^{-{3\over 4} } g_6^{1\over 2} \ell_s\,,\quad L_{{\rm Planck},6}=(4\pi^3)^{1\over 4}   g_6^{1\over 2} \ell_s\,,
\fe
in terms of the string scale $\ell_s$ and the 6d string coupling defined by
\ie 
g_6^2= {g_s^2 (2\pi \ell_s)^4\over V_M}\,,
\label{g6}
\fe
where $g_s$ is the type IIB string coupling and $V_M$ is the volume of the internal four-manifold $M$ in the string frame. The appropriate flat space limit here amounts to 
taking $L_{\rm AdS}/\ell_s \to \infty$ and keeping $g_6$ fixed.

The moduli direction saturating the lower bound in \eqref{AdSgravboundSUSY} amounts to the large radius limit of the seed CFT which is universal to the symmetric orbifold CFTs (see Section~\ref{sec:limitingCFT} for the general discussion). Here it maps to the large radius limit in the target space $M$. Therefore, the light operators approaching vanishing gap in the infinite distance limit of the CFT naturally correspond to Kaluza-Klein (KK) modes on $M$ with vanishing mass in this limit. To identify the mass decay rate of such KK modes, we start with the IIB action in the Einstein frame \cite{Polchinski:1998rq},
\ie 
S_{\rm IIB}={1\over 2 \kappa^2} \int d^{10} x\sqrt{g}\left( R-{1\over 2} (\pa \Phi)^2+\dots \right)
\fe
with the standard convention $g_s=e^\Phi$ for the dilaton and $\kappa=L_{{\rm Planck},10}^4$, and perform KK reduction 
on the product metric
\ie 
ds^2= e^{-\rho} ds_{6}^2+ e^{\rho} ds_M^2\,,
\label{IIBmetric}
\fe
where $\rho$ is a 6d scalar encoding the radius modulus of the internal manifold $M$. The resulting 6d action including the kinetic terms for the moduli scalars $\Phi$ and $\rho$ reads
\ie 
S_{6d}={1\over 2 \kappa_6^2} \int d^{6} x\sqrt{g_6}\left( R_6-{1\over 2} (\pa \Phi)^2 
-2(\pa\rho)^2+
\dots \right)
\fe
where $\kappa_6=L_{{\rm Planck},6}^2$. Now the KK mass for large $M$, in the 6d Planck units, is
\ie 
m_{\rm KK} \sim e^{-\rho}\,,
\label{mKKdecay}
\fe
where one factor of $e^{-{\rho\over 2}}$ comes from the inverse radius of $M$ in \eqref{IIBmetric} and the other factor of $e^{-{\rho\over 2}}$ comes from the difference between the 10d and 6d Planck lengths due to the internal volume. In terms of the canonically normalized scalar $\hat \rho=\sqrt{2}\rho$ (in 6d Planck units), the mass decays as $m_{\rm KK} \sim e^{-\hat \rho/\sqrt{2}}$ as in  \cite{Etheredge:2022opl} and appears faster than predicted by the CFT in \eqref{flatlowerboundSUSY}.

However the particular direction towards infinite distance as inherited by taking the flat space limit of the AdS result must involve varying dilaton $\Phi$. Indeed, it follows from \eqref{g6} that the 6d string coupling depends on the moduli fields,
\ie 
g_6\propto e^{\Phi-2\rho}\,,
\fe
where we have used $V_M\propto e^{\Phi+2\rho}$ in the string frame. Consequently to hold $g_6$ fixed in this procedure requires the following identification between the moduli fields $\Phi,\rho$ and the 
 canonically normalized scalar $\phi$ in \eqref{canscalar} inherited from the theory in AdS,
 \ie 
\Phi=\phi\,,\quad \rho={1\over 2}\phi\,.
 \fe
 It then follows from \eqref{mKKdecay} that
 \ie 
 m_{\rm KK} \sim e^{-{1\over 2}\phi}\,,
\fe
as predicted by \eqref{AdSgravboundSUSY}, in agreement with \cite{Etheredge:2022opl}.\footnote{As explained in \cite{Etheredge:2022opl}, this same decay rate is also saturated by the string oscillator modes.} 

It would be interesting to generalize this analysis to more general CFTs and also to higher dimensions, and compare with the conjectured flat space bounds.

\subsection{Organization of This Paper}

This paper is organized as follows. In Section~\ref{sec:examples}, we discuss examples of singularities of conformal manifolds. There are singularities at finite distances, where $\Delta_{\rm gap}$ remains non-zero, and there are singularities at infinite distance (also known as cusp points), where $\Delta_{\rm gap}$ vanishes. We also present examples with the decay rate saturating the upper bound $\alpha =1$, where all marginal operators are exact  in the limit, and with $\alpha < 1$, where there could be marginal operators that are not exact in the limit. 
In Section~\ref{sec:zerogaptoinfdist}, we prove Theorems~\ref{thm:zerogaptoinfdist} and~\ref{thm:gapvanishing} using conformal bootstrap for the
four-point function of the operator with vanishing conformal dimension. In Section~\ref{sec:limitingCFT}, we prove Theorems~\ref{thm:CFTinthelimit} and~\ref{thm:lowerbound}. We end with discussion on future directions in Section~\ref{sec:discussion}.

\section{Examples of Conformal Manifold in $d=2$}
\label{sec:examples}

To illustrate the theorems we prove in concrete terms, let us discuss in more detail three examples of conformal manifolds in $d=2$ CFTs.

\subsection{Narain Moduli Space of $c=2$ Toroidal CFT}
The first example we consider is the $c=2$ toroidal CFT defined by a sigma-model on the two-torus $T^2=\mR^2/\Lambda$ for integral lattice $\Lambda\equiv \mZ e_1\oplus \mZ e_2$ and target coordinates $X\equiv X^1 e_1 +X^2 e_2\in \mR^2$. The CFT action reads\footnote{We follow the convention of \cite{Polchinski:1998rq} with $\A'=2$.}
\ie 
S={1\over 4\pi }\int d^2 z\, (G_{ij} + B_{ij})\pa X^i \bar \pa X^j\,,
\label{t2sigmamodel}
\fe
where $G_{ij}\equiv e_i \cdot e_j$ is the metric on the $T^2$ and $B_{ij}=-B_{ji}$  is the B-field.

The full $c=2$ conformal moduli space has a complicated branch structure \cite{Dulat:2000xj}. Here we focus on the conformal submanifold that parameterizes geometric deformations of \eqref{t2sigmamodel}, namely those generated by the four exactly marginal operators $\pa X^i \bar\pa X^j$ which change  $G_{ij},B_{ij}$ in \eqref{t2sigmamodel}. This conformal submanifold is known as the Narain moduli space, 
\ie 
\cM_{\rm Narain}={\rm O}(2,2;\mZ)\backslash{\rm O}(2,2;\mR)/{\rm O}(2)\times {\rm O}(2)\,,
\label{narain}
\fe 
which is an orbifold of a symmetric space of real dimension 4 and the left quotient by ${\rm O}(2,2;\mZ)$ implements the identifications due to duality transformations.  
The local primary operators with respect to the $\mf{u}(1)^2$ current algebra
\ie 
\cO_{ m, w}=e^{i  p_L(m,w) \cdot X_L +i  p_R(m,w) \cdot X_R }\,,\quad h={ p_L\cdot p_L \over 2}\,,\quad \bar h={ p_R\cdot p_R\over 2}\,,
\label{torusops}
\fe
are parametrized by momentum and winding vectors $m\in \Lambda$ and $w\in \Lambda^*$ respectively where $\Lambda^*$ denotes the dual lattice and their left and right conformal weights are also listed above. Here $X_L,X_R$ denote the left and right components of the scalar field $X$ and 
$p_L,p_R$ are the left and right  momenta measured by the corresponding $\mf{u}(1)$ symmetries. The pair $(p_L,p_R)$ defines an embedding of the charge lattice $\Lambda\oplus \Lambda^*$ as an even self-dual lattice in $\mR^{2,2}$ (i.e $p_L$ and $p_R$ live in the positive and negative $\mR^2$ subspaces respectively), as required by locality and modular invariance of the CFT.  The Narain moduli space \eqref{narain} then naturally parametrizes such embeddings up to automorphisms of the charge lattice.  

There is another representation of the Narain moduli space \eqref{narain} which is more convenient for the $c=2$ case here \cite{Dulat:2000xj},
\ie 
\cM_{\rm Narain}
=
\left ({\mH_\sigma   \over {\rm PSL}(2,\mZ)} \times {\mH_\rho \over {\rm PSL}(2,\mZ)} \right)/\mZ_2\times \mZ_2\,,
\label{Narainc2}
\fe
where $\sigma=\sigma_1+i\sigma_2$ is the complex structure moduli and $\rho=\rho_1+i\rho_2$ is the complexified K\"ahler moduli for the target $T^2$, both taking values in the upper half plane $\mH$. The duality groups ${\rm PSL}(2,\mZ)$ act on $\sigma$ and $\rho$ respectively. The residual $\mZ_2\times \mZ_2$ corresponds to a swap (mirror) $(\sigma,\rho) \to (\rho,\sigma)$ and a reflection $(\sigma,\rho) \to (-\bar\sigma,-\bar\rho)$. In this parametrization, the left and right conformal weights in \eqref{torusops} are
\ie 
h_L={1\over 4 \rho_2\sigma_2}|m_2-m_1\sigma-\rho(w_1+w_2\sigma)|^2\,,\quad h_R={1\over 4 \rho_2\sigma_2}|m_2-m_1\bar\sigma-\rho(w_1+w_2\bar\sigma)|^2\,,
\label{torusweights}
\fe
where $m_1,m_2,w_1,w_2\in \mZ$ denote the momentum and winding charges. 

The Zamolodchikov metric on the Narain moduli space respects the symmetric space structure \eqref{narain} and in terms of \eqref{Narainc2} it follows from the Poincare metric on the upper half plane,
\ie 
ds^2={2 d\sigma d\bar \sigma \over \sigma_2^2} + {2 d\rho d\bar \rho \over \rho_2^2}\,.
\fe
We restrict $\sigma,\rho$ to their standard fundamental domains $\cF\subset\mH$, which are subject to further discrete identifications by $\mZ_2\times \mZ_2$ in \eqref{Narainc2}. The overall normalization of the metric above is fixed by \eqref{Sdef}.
More explicitly, for rectangular torus, $\rho_1=\sigma_1=0$, $\rho_2={R_1 R_2\over 2}$ and $\sigma_2={R_1\over R_2}$, the metric reduces to 
\ie 
ds^2=  \sum_{i=1}^2 {4\over R_i^2}(dR_i)^2 \,,
\fe 
where each summand is the Zamolodchikov metric for the $S^1$ sigma-model \cite{Moore:2015bba} in our normalization.\footnote{Note that it follows from \eqref{Sdef} that our normalization of the Zamolodchikov metric for 2d CFT differs from that of \cite{Moore:2015bba} by $ds^2_{\rm here}=(2\pi)^2 ds^2_{\rm there}$.}

The infinite distance points on $\cM_{\rm Narain}$ are located at the cusp $\rho=i\infty$ (for any $\sigma$) up to a choice of the duality frame, where the shortest geodesic distance to any point $(\sigma^*,\rho^*)$ in the interior of $\cF$ is 
\ie 
t=\sqrt{2}\left(\log^2 {\rho_2\over \rho^*_2}+\log^2 {\sigma_2\over \sigma^*_2}\right)^{1\over 2}
\label{toroidaldist}
\fe 
which diverges as 
\ie 
t= \sqrt{2}\log \rho_2 + {\rm finite}\,,
\label{toroidaldist1}
\fe
as $\rho_2\to \infty$ when $\sigma_2$ is fixed, or 
\ie 
t= \log \rho_2 \sigma_2 + {\rm finite}\,,
\label{toroidaldist2}
\fe
if $\rho_2={\gamma^2\over 2}\sigma_2\to \infty$ for $\C>0$.

It's immediate from \eqref{torusweights} that an infinite tower of operators with zero winding charge $w_1=w_2=0$ obtain vanishing conformal weights at infinite distance as stated in Theorem~\ref{thm:zerogaptoinfdist}. The bottom of the tower has $m_1=1,m_2=0$ and gives the dimension gap,
\ie 
\lim_{\rho \to i\infty}\Delta_{\rm gap}(\sigma,\rho)={1\over 2\rho_2\sigma_2}
\fe
Comparison with \eqref{toroidaldist} confirms the exponential approach to vanishing gap with rate $\A=1$ for the limit \eqref{toroidaldist2} and $\A={1\over \sqrt{2}}$ for the limit \eqref{toroidaldist1}, in accordance with Theorem~\ref{thm:gapvanishing} and Theorem~\ref{thm:lowerbound}, saturating the bounds thereof. Note that Theorem~\ref{thm:CFTinthelimit} is confirmed tautologically in this case:
in the limit \eqref{toroidaldist2}, the CFT is described by the $T^2$ sigma-model with large radius and in the limit \eqref{toroidaldist1}, the CFT is described by a $S^1$ sigma-model with large radius and another $S^1$ sigma-model of finite radius $R=\C$.

The other singularities on the conformal manifold $\cM_{\rm Narain}$ are of the orbifold type and locate at finite distance, corresponding to either the $\mZ_2$ fixed point at $\rho=i$ or the $\mZ_3$ fixed point $\rho=e^{2\pi i\over 3}$ on $\cH_\rho$ and similarly for $\cH_\sigma$. In particular, the maximal dimension gap $\Delta_{\rm gap}={2\over 3}$ is achieved at the simultaneous $\mZ_3$ fixed point $\rho=\sigma=e^{2\pi i\over 3}$, as can be seen by inspecting \eqref{torusweights}. This point on $\cM_{\rm Narain}$ is described by the $SU(3)_1$ Wess-Zumino-Witten CFT and the gap is saturated by the nontrivial Kac-Moody primary operators.

\subsection{K\"ahler Moduli Space of $c=6$ Quintic CFT}
\label{sec:quinticexample}

The second example we consider is the $\cN=(2,2)$ SCFT defined by a supersymmetric sigma-model with target space defined by the quintic Calabi-Yau manifold realized as a hypersurface $W$ of degree 5 in $\mP^4$. The CFT action again takes the form as in \eqref{t2sigmamodel} 
with additional fermion fields that furnish the supersymmetric completion.
The quintic SCFT has a large conformal manifold parametrized by 101 complex structure moduli and 1 complexified K\"ahler structure moduli $\tau \in \mH$, which encode exactly marginal deformations of the target space metric and B-field. Here we focus on the complex 1-dimensional submanifold $\cM_{\rm K\ddot{a}hler}(W)$ parametrized by the K\"ahler moduli $\tau$, which couples to the exactly marginal operator $\omega_{i \bar j} \pa X^i \bar\pa X^{\bar j}$ corresponding to the harmonic $(1,1)$-form on $W$ where $X^i,X^{\bar j}$ with $i,\bar j=1,2,3$ are complex coordinates for $W$.

By mirror symmetry, $\cM_{\rm K\ddot{a}hler}(W)$ is equivalent to the complex structure moduli of the mirror quintic $\widehat W$, defined as the hypersurface orbifold,
\ie 
\{[Z_a]\in \mP^4 \,|\, \sum_{a=1}^5 Z_a^5  -5 \psi \prod_{a=1}^5Z_a =0\}/ \mZ_5^3\,,
\fe
where the $\mZ_5^3$ are generated by rotations $Z_a\to e^{2\pi i n_a\over 5}Z_a$ on the homogeneous coordinates of $\mP^4$ preserving the polynomial equation that defines the hypersurface. The orbifold restricts possible complex structure deformations of the hypersurface, and the complex parameter $\psi$ is the unique moduli that survives. Moreover, redefinition of the homogeneous coordinates $Z_a$ ($e.g.$ $Z_1 \to e^{2\pi i\over 5} Z_1$) induces an identification on $\psi$ such that the true moduli space is 
\ie 
\cM_{\rm cs}(\widehat W)=\cM_{\rm K\ddot{a}hler}(W)=\{\psi \in \mC\,|\,\psi \sim e^{2\pi i \over 5} \psi \}\,,
\label{quinticKahler}
\fe
where the relation between the K\"ahler moduli $\tau$ and the complex structure moduli $\psi$ are related by the mirror map \cite{Candelas:1990rm}.
Equivalently, we work with a fundamental domain parametrized by $0\leq \arg \psi <{2\pi \over 5}$.

The Zamolodchikov metric on the moduli space \eqref{quinticKahler}, up to an overall normalization, is equal to the standard Weil-Petersson metric which is determined by the special geometry relations in terms of the period integrals on the mirror quintic $\widehat W$ \cite{Candelas:1989qn,Candelas:1990rm}.
Let us now summarize the singularity structures on $\cM_{\rm cs}(\widehat W)$ (equivalently $\cM_{\rm K\ddot{a}hler}(W)$).

There are three singularities on $\cM_{\rm cs}(\widehat W)$, which has the topology of a three-punctured sphere. The three singularities are of different natures.

\subsubsection*{Orbifold Singularity}

Firstly, there is an orbifold singularity at $\psi=0$ which is fixed by the $\mZ_5$ identification in \eqref{quinticKahler}. This is known as the Gepner point on $\cM_{\rm cs}(\widehat W)$, where the CFT is completely regular and described by a $\mZ_5^3$ orbifold of the $\cN=2$ Landau-Ginzburg model with superpotential $\sum_{a=1}^5 Z_a^5$, equivalently a tensor product of five copies of the Kazama-Suzuki $\cN=2$ supercoset model $SU(2)_3/U(1)$. The dimension gap at the Gepner point is,
\ie 
\Delta_{\rm gap}^{\rm Gepner}={2\over 5}\,,
\label{gepnergap}
\fe
which is saturated by a non-BPS primary operator of zero $U(1)_R$ charge that arises from a product of BPS and anti-BPS primaries of dimension $\Delta={1\over 5}$ in two copies of the supercoset model.

\subsubsection*{Conifold Singularity}

Secondly, there is a curvature singularity on the moduli space located at $\psi=1$, known as the conifold point on $\cM_{\rm cs}(\widehat W)$. In terms of the K\"ahler moduli of the quintic CY, this corresponds to a point of purely imaginary $\tau$.
The CFT at the conifold point is singular and develops a continuous spectrum above a nonzero gap
\ie 
\Delta_{\rm gap}^{\rm Conifold}={1\over 2}\,.
\label{conifoldgap}
\fe
The continuum is described by the $\cN=2$ Liouville CFT of central charge $c=9$ \cite{Mukhi:1993zb,Ooguri:1995wj}, which is equivalent to the $\cN=2$ cigar CFT defined by the Kazama-Suzuki supercoset $SL(2)_1/U(1)$ \cite{Kazama:1988qp}, and the dimension gap \eqref{conifoldgap} is saturated by the supercoset primary $\Phi_{j,m,\bar m}$ with $j=-{1\over 2}$ and $m=\bar m=0$.\footnote{The $\cN=2$ Liouville (cigar) operators are normalized in a different way than those of the canonical normalization in the compact CFT, due to the divergent volume factor in the non-compact Liouville direction. Consequently, finite correlation functions in the $\cN=2$ Liouville CFT translate into divergent correlation functions of normalized operators in the quintic SCFT at the conifold point: the simplest example being the chiral ring coefficient which measures the three-point function of (anti)chiral ring operators. See related discussions in \cite{Lin:2015wcg,Lin:2016gcl} for the four-point functions of (anti)chiral ring operators.} 
Let us parametrize the region near the conifold point in polar coordinates by $\psi=1+r e^{i\theta}$. The Zamolodchikov metric and its Ricci curvature scalar $R$ are locally given by the following \cite{Candelas:1990rm}
\ie 
ds^2=-a^2 \log r (dr^2+r^2 d\theta^2)\,,\quad R={1\over 2 a r^2 (-\log r)^2}\,, \quad r\ll1\,,
\fe
where $a>0$ is a constant. Note that despite the curvature singularity, the conifold point is at finite distance on $\cM_{\rm cs}(\widehat W)$, which is consistent with our Theorem~\ref{thm:zerogaptoinfdist} and the non-vanishing gap \eqref{conifoldgap}.\footnote{The fact that conifold points reside at finite distance on the moduli space of Calabi-Yau manifolds holds in general 
\cite{Candelas:1988di,Candelas:1989ug} and they are portals to connecting Calabi-Yau manifolds of different topology via  geometric transition \cite{Candelas:1989ug,Candelas:1989js,Strominger:1995cz}. Indeed, it was conjectured by Reid \cite{reid1987moduli} (also known as Reid's fantasy) that all Calabi-Yau three-folds are connected this way and there is a complete universal moduli space for all 
(see also \cite{Green:1988bp}). The topology change comes from shrinking a two-cycle of the resolved conifold and expanding a three-cycle in the deformed conifold (and vice versa) and this transition is smooth in the full string theory by including branes wrapping the vanishing cycles \cite{Strominger:1995cz}. It would be interesting to understand the corresponding connection in the space of 2d $\cN=(2,2)$ supersymmetric QFTs and prove the Reid conjecture using QFT methods.}

\subsubsection*{Large Complex Structure Limit}

Finally, the remaining singularity on $\cM_{\rm cs}(\widehat W)$ comes from the large complex structure limit $\psi=\infty$ of the  mirror quintic $\widehat W$, which is equivalent to the large volume limit  $\tau=i\infty$ of the quintic. The mirror map that relates the two take the following form in this limit \cite{Candelas:1990rm},
\ie 
\tau ={5i\over 2\pi} \log \psi +{\rm 
finite}\,.
\label{lvmirrormap}
\fe
The Zamolodchikov metric and associated scalar curvature in this limit are\footnote{Note that our normalization of the Zamolodchikov metric (which follows from \eqref{Sdef}) differs from that of \cite{Candelas:1989qn}  by $ds^2_{\rm here}=4 ds^2_{\rm there}$.} 
\ie 
ds^2={6\over  |\psi|^2 \log^2 |\psi|} d\psi d\bar\psi={6\over \tau_2^2}d \tau d\bar\tau\,,\quad R= -{1\over 3}\,,\quad |\psi|,\tau_2\gg 1\,,
\fe
where the change of coordinates between $\psi$ and $\tau$ comes from \eqref{lvmirrormap}. The sigma-model on the quintic CY $W$ in the large volume limit has an infinite tower of low lying states coming from eigenfunctions of the scalar Laplacian on $W$ and the eigenvalues correspond to their scaling dimensions. By dimensional analysis, we conclude that the dimension gap depends on the diameter $L$ of $W$ as follows,
 \ie 
\Delta_{\rm gap}\sim {1\over L^2}\sim {1\over \tau_2}\,,
 \fe 
 where in the last step we have used fact that $\tau_2$ measures the integral of the K\"ahler class. On the other hand, the geodesic distance to the infinite distance limit $\tau_2\to \infty$ diverges as
 \ie 
t=\sqrt{6}\log \tau_2 +{\rm finite}\,.
 \fe
 Consequently, we find that $\Delta_{\rm gap}$ vanishes exponentially in this limit with rate $\A={\sqrt{1\over 6}}$, in accordance with Theorems~\ref{thm:zerogaptoinfdist},~\ref{thm:gapvanishing} and the prediction for $\A$ from Theorems~\ref{thm:CFTinthelimit} and~\ref{thm:lowerbound}. In particular, the compact CFT factor in this infinite distance limit has $c=3$ and is described by the six fermions necessary for the $\cN=(2,2)$ supersymmetry. The generalization of the above discussion to more general $\cN=(2,2)$ superconformal sigma-models (including orbifolds) are straightforward.

\subsection{Non-unitary Counterexample}
As emphasized in the introduction, the general results in this work apply to unitary CFTs. Here to illustrate the importance of unitarity, let us describe a simple example of a non-unitary CFT with conformal manifold where the Theorems~\ref{thm:zerogaptoinfdist},~\ref{thm:gapvanishing},~\ref{thm:CFTinthelimit}, and~\ref{thm:lowerbound} are no longer applicable. In particular, we will see that in the non-unitary theory, the limit of vanishing gap (scaling dimensions) on the conformal manifold can happen at finite distance with respect to the Zamolodchikov metric. 

The model we consider is a simple modification of the previous example.
We take the tensor product CFT $\cT_{c={15\over 4}}$ of the quintic SCFT with $c=9$ and a non-unitary minimal model $M_{3,8}$ of central charge $c=-{21\over 4}$. The $M_{3,8}$ CFT contains, among other scalar operators, the identity operator and a scalar primary operator of the lowest scaling dimension $\Delta=-{1\over 2}$. This product theory $\cT_{c={15\over 4}}$ thus has a normalizable identity operator,  a conformal manifold that coincides with that of the quintic SCFT, and a discrete operator spectrum at generic points on the conformal manifold. As in Section~\ref{sec:quinticexample}, we focus on the conformal submanifold $\cM_{\rm K\ddot{a}hler}(W)$.

By taking products with the $\Delta=-{1\over 2}$ operator in the $\cM_{3,8}$ CFT, the continuum starting at $\Delta={1\over 2}$ 
(see around
\eqref{conifoldgap}) at the conifold point of the quintic SCFT is brought down to $\Delta=0$, producing a continuum above the identity operator in the product theory $\cT_{c={15\over 4}}$. Since the conifold point is at finite distance on $\cM_{\rm K\ddot{a}hler}(W)$, this clearly creates a counterexample for our theorems in the case of non-unitary CFTs. Said differently, near the Gepner point of $\cM_{\rm K\ddot{a}hler}(W)$ with dimension gap \eqref{gepnergap} in the quintic SCFT, the product CFT $\cT_{c={15\over 4}}$ has an almost non-negative operator spectrum except for a few low-lying states. The non-unitarity becomes much more severe as one wanders around on the conformal manifold, in particular after passing the conifold point where $\Delta_{\rm gap}=0$ and towards the large volume limit where infinitely many non-unitary operators appear. This is possible because $\Delta_{\rm gap}=0$ happens at finite distance in this non-unitary theory. It may be interesting to formulate a modified degeneration limit for non-unitary CFTs ($e.g.$ accumulation in the spectrum to the lowest state)
and correspondingly a version of
our theorems that would apply to the non-unitary context but that is beyond the scope of this work.

\section{From Vanishing Gap to Infinite Distance}
\label{sec:zerogaptoinfdist}

In this section, we study the limits on the conformal manifold $\cM$ where the conformal dimension gap $\Delta_{\rm gap}$ vanishes, and prove that they are at infinite distance from any interior points of $\cM$ measured with respect to the Zamolodchikov metric. More precisely, for any geodesic $\lambda(t)$ on $\cM$ parametrized by proper distance $t$ and any finite $T>0$, there exists $\ve>0$ such that $\Delta_{\rm gap}(t)\equiv \Delta_{\rm gap}(\lambda(t))\geq \ve$ for $t\in [-T,T]$.  We assume that the four-point functions of light primary fields are well-defined in the limit. In particular we assume up to rescaling there is a unique operator of conformal weights $h=\bar h=0$ which coincides with the identity operator.

We will make use of the following formula from conformal perturbation theory \cite{Zamolodchikov:1987ti}, 
\ie 
 \frac{d \Delta(t)}{dt} = - C_{{\cal O}{\cal O} M}(t)\,,
    \label{ST}
    \fe 
which determines how the scaling dimension $\Delta(t)$ of a hermitian operator $\cO$ changes along the geodesic parametrized by $t$, in terms of the OPE coefficient with the exactly marginal operator $M$ that couples to $t$. Note that both $\cO$ and $M$ are normalized to have unit two-point functions.

We proceed by contradiction. We assume that there exists finite $t_*>0$ along a geodesic $\lambda(t)$ such that  $\Delta(t=t_*)=0$. We will show that this is incompatible with \eqref{ST} using conformal symmetry and crossing invariance of the CFT. In particular, we note that if the following limit exists,
\ie
\lim_{t\to t_*} {C_{\cO\cO M}(t)\over \Delta(t)}=\alpha\,,
\label{ratioCD}
\fe 
then by integrating \eqref{ST} for $t-t_*\ll t_*$, we have
\ie 
\log \Delta(t) -\log \Delta(t_*) \sim -\alpha (t-t_*)\,.
\fe
However this is clearly impossible if $\Delta(t_*)$ is vanishing since the left-hand side will diverge. Therefore it suffices to derive \eqref{ratioCD} for a scalar operator ${\cal O}$ whose scaling dimension $\Delta(t_*)$ can be made arbitrarily small and such $\cO$ exists by assumption.\footnote{By the  von Neumann–Wigner non-crossing theorem (see \cite{Korchemsky:2015cyx} for a CFT-related 
discussion), level-crossing on the conformal manifold should only appear in higher codimensions. Hence close to the limit of vanishing gap we can choose a geodesic 
 transverse to the level-crossing loci.}
In the following, to ease the notations, all the CFT quantities are assumed to be evaluated closed to $t=t_*$ and we will simply denote $\Delta(t_*)$ by $\Delta$.

It follows from the $d=2$ global conformal algebra $\mf{sl}(2,\mR)\times \mf{sl}(2,\mR)$ that the left and right conformal descendants of $\cO$ satisfy,
\ie 
    \partial {\cal O} = i \sqrt{\Delta} J\,,
\quad 
 \bar \partial {\cal O} = i \sqrt{\Delta}  \bar J\,,\quad \pa  \bar J = \bar\pa J=i\sqrt{\Delta} K\,,
 \label{OtoJJb}
\fe 
for a triplet hermitian operators $J,\bar J,K$ of conformal weights $(\frac{1}{2} \Delta+1, \frac{1}{2}\Delta)$, $(\frac{1}{2} \Delta, \frac{1}{2}\Delta+1)$ and $(\frac{1}{2} \Delta+1, \frac{1}{2}\Delta+1)$ respectively, with unit-normalized two-point functions.  This can also be verified by taking derivatives 
of the two-point function
\ie 
    \langle {\cal O}(z) {\cal O}(w) \rangle 
    = \frac{1}{|z-w|^{2\Delta}}\,,
    \fe
which upon acting with $\pa_z \pa_w$ and $\bar\pa_{\bar z} \bar\pa_{\bar w}$ gives 
\begin{equation}
    \langle J(z) J(w) \rangle =
    \frac{1}{(z-w)^2} + O(\Delta)\,,\quad 
     \langle \bar J(z) \bar J(w) \rangle =
    \frac{1}{(\bar z-\bar w)^2} + O(\Delta)\,,
    \label{JJ}
\end{equation}
and from $\pa_z \bar\pa_{\bar w}$
\begin{equation}
    \langle J(z) \bar J (w) \rangle = 
    \frac{\Delta}{|z-w|^2} + O(\Delta^2)\,,
    \label{JJmixed}
\end{equation}
and similarly for the correlation functions involving the scalar operator $K$.  

From the above, though $J,\bar J$ and $K$ are descendants of ${\cal O}$,  they behave in the limit $\Delta \to 0$ as primary operators of conformal weights $(1,0),(0,1)$  and $(1,1)$ respectively. Naturally, they arise from the decomposition of the generic global conformal multiplet approaching the unitarity bound.

Let us now consider the three-point function,
\begin{equation}
\langle {\cal O}(z){\cal O}(w)
M(u) \rangle 
= \frac{C_{{\cal O}{\cal O} M}}{(z-w)^{\Delta -1} (z-u) (w-u) (\bar z - \bar w)^{\Delta -1} (\bar z - \bar u)(\bar w - \bar u)}\,.
    \label{OOM}
\end{equation} 
By acting $\partial_z\bar \partial_{\bar w}$ on both sides of \eqref{OOM} and using \eqref{OtoJJb}, we obtain the following relation between the OPE coefficients involving $\cO$ and its normalized descendants $J,\bar J$,
\begin{equation}
    C_{{\cal O}{\cal O} M}
          =  \Delta C_{J\bar J M}
          \left( 1 + O(\Delta) \right)\,.
          \label{JbarJMOPE}
\end{equation}
Therefore, $\alpha=C_{J\bar J M}$ in \eqref{ratioCD} and it suffices to show that $|C_{J\bar J M}|<\infty$.
In fact, we are going to derive a stronger statement that 
\begin{equation}
{C_{J\bar J M}} \leq  1 + O(\Delta)\,,
\label{CJJMbound}
\end{equation}
from the crossing invariance of the four-point function of the scalar operator $\cO$ whose dimension $\Delta$ can be made arbitrarily close to zero. We will also identify the necessary and sufficient conditions to saturate the upper bound on ${C_{J\bar J M}}$.

In this paper, we do not assume that the CFT satisfies the CFT axioms in the  $\Delta \rightarrow 0$ limit. For example, the genus-one partition function may diverge in the limit. What we assume is that the CFTs satisfies the axioms {\it before} we take the limit and the four-point functions of the light operators are well-defined in the limit. 
If we allow ourselves to examine the four-point functions of $J$ and $\bar J$ directly at the limit, we would have a simple derivation of \eqref{CJJMbound}. Therefore, we will first present this simple but not rigorous derivation of \eqref{CJJMbound} in order to provide an intuitive understanding of it. We will then give a more rigorous proof without making this additional assumption. 

Here is the non-rigorous derivation of \eqref{CJJMbound}. Since $J(z)$ and $\bar J(\bar z)$ are primary fields of conformal weights $(1,0)$ and $(0,1)$ in the limit, the four-point function 
$\langle  \bar J ( \bar w) J(z)\bar J(\bar u)   J(  v)\rangle  $ is holomorphic in $z, v$ and anti-holomorphic in $\bar w, \bar u$ as $\Delta \to 0$. By \eqref{JJ}, it has the $t$-channel expansion as,
\ie
\langle  \bar J ( \bar w) J(z)\bar J(\bar u)   J(  v)\rangle  = \frac{1}{(z-v)^2(\bar w - \bar u)^2} +\dots \,.
\label{nonrigorousone}
\fe
On the other hand, in the $s$-channel, it can be expanded as,
\ie
\langle  \bar J ( \bar w) J(z)\bar J(\bar u)   J(  v)\rangle  =\frac{G^{ij}  C_{J \bar J M_i} C_{J \bar J M_j}}{(z-v)^2 (\bar w - \bar u)^2} + \cdots,
\label{nonrigoroustwo}
\fe
where we introduced a hermitian basis of such operators denoted by $M_i$ for $i=1,2,\dots$ and define a metric $G_{ij}$ by
\ie 
\la M_i(z) M_j(w)\ra ={G_{ij}\over |z-w|^4}\,.
\fe
A subset of them are exactly marginal which we choose to be $M_a$ with $a=1,2,\dots,\dim \cM$ and $G_{ab}$ for them is the Zamolodchikov metric, and  $M_i$ with $i > \dim \cM$ are not exactly marginal if they exist. 

It turns out that the global conformal block for the identity exchange in the $t$-channel and that for the marginal operators exchange in the $s$-channel take the identical form of 
$(z-v)^{-2}(\bar w - \bar u)^{-2}$ in \eqref{nonrigorousone} and \eqref{nonrigoroustwo} respectively.\footnote{In fact in both cases it captures the entire Virasoro block for the given set of internal and external weights.}
Contributions of other operators indicated by 
$(\cdots)$ in these equations have different functional dependence on $\zeta$ and are suppressed in the $\Delta \rightarrow 0$ limit. 
Comparing \eqref{nonrigorousone} and \eqref{nonrigoroustwo}, it follows from the associativity of the OPE directly in the limit that 
\ie
G^{ij}  C_{J \bar J M_i} C_{J \bar J M_j} = 1 \, .
\fe
Since the exactly marginal operator of interest $M$  is a linear combination of $M_i$ and has a unit two-point function, we deduce that
\ie
C_{J\bar JM} \leq 1\,.
\fe
This derivation is not rigorous since we have used unitarity and associativity of OPE for $J$ and $\bar J$  directly in the limit, despite the fact the limiting CFT does not obey the usual axioms of compact unitary CFT. In particular, the fact that the ground state becomes non-normalizable with vanishing gap in the limit raises a concern.\footnote{For example, in such a CFT, vanishing right (resp. left) conformal weight does not necessarily imply holomorphicity (resp. antiholomorphicity) of the operator. A typical example is the $SO(n)$ rotation symmetry currents $X^{[i}\pa_\m X^{j]}$ for $n$ non-compact bosons $X^i$ with $i=1,\dots,n$, whose components are neither holomorphic nor anti-holomorphic. See \cite{Dubovsky:2023lza} for a recent discussion on such currents and how they arise from the vanishing gap limit of Wess-Zumino-Witten CFT as the level $k\to \infty$.}
We have also assumed that the identity and marginal operator exchanges in \eqref{nonrigorousone} and \eqref{nonrigoroustwo}, respectively, are linearly independent of other terms in the expansions indicated by $(\cdots)$, and this requires a proof. 

In the following we present a more rigorous derivation of the same inequality.
The four-point function of ${\cal O}$'s can be expanded into a sum over $s$-channel global conformal blocks as below,
\ie 
\langle {\cal O}(w)  {\cal O}(z)
 {\cal O} (u)  {\cal O}(v) \rangle
 = \frac{1 + \sum_{\phi\neq 1} |C_{{\cal O}{\cal O}\phi}|^2
 {\cal F}_{h_\phi}(\zeta)\bar {\cal F}_{\bar h_\phi}(\bar\zeta) }
{|z-w|^{2\Delta} |u-v|^{2\Delta}}\,,
\label{O4pf}
\fe 
where we have separated the contribution of the identity operator from those of the nontrivial normalized global primaries $\phi$ with  conformal weights $(h_\phi, \bar h_\phi)$. The holomorphic conformal cross-ratio $\zeta$ is defined as
\begin{equation}
    \zeta=\frac{(z-w)(u-v)}{(z-u)(w-v)}\,,
\end{equation}
and similarly for the antiholomorphic cross-ratio $\bar\zeta$. The corresponding global conformal blocks are denoted as 
${\cal F}_{h_\phi}(\zeta) $  and $\bar {\cal F}_{\bar h_\phi}(\bar\zeta) $, which are independent of $\Delta$ and explicitly given by,
\ie 
{\cal F}_{h_\phi}(\zeta)= \zeta^{h_\phi} {}_2 F_1(h_\phi, h_\phi, 2h_\phi; \zeta)\,.
\label{globalblock}
\fe 
Taking the derivatives $\partial_z \bar \partial_{\bar w} \bar\partial_{\bar u} \partial_{v}$ on
the two sides of \eqref{O4pf}
and using the relation \eqref{OtoJJb}, we obtain
\ie 
&\langle  \bar J (  w) J(z)\bar J(u)
  J(  v)\rangle  =\sum_{\phi\neq 1}
\frac{|C_{{\cal O}{\cal O}\phi}|^2}{\Delta^2}\
\frac{\partial^2 {\cal F}_{h_\phi}(\zeta)}{\partial z \partial v}\
\frac{\partial^2 \bar {\cal F}_{\bar h_\phi}(\bar\zeta)}{\partial \bar w \partial \bar u} + O(\Delta)\,,
 \label{J4pfs}
\fe 
where we have dropped the identity contribution in the $s$-channel OPE since it is of order $O(\Delta^2)$ from 
\eqref{JJmixed}.

Let us investigate the contributions from marginal operators (which include $M$) on the RHS of \eqref{J4pfs}. 
 Using the explicit form of the global conformal block for marginal operators,
\ie 
{\cal F}_1(\zeta)={\zeta} \,{}_2 F_1(1, 1, 2; \zeta) =
    - \log(1-\zeta)\,,
    \label{confblockmarginal}
\fe
and the relation \eqref{JbarJMOPE}, which holds for general marginal operators in place of $M$, we find that $M_i$ contributes to the RHS of \eqref{J4pfs} by the following,
\ie 
G^{ij}  C_{J \bar J M_i} C_{J \bar J M_j} \
\frac{\partial^2 {\cal F}_1(\zeta)}{\partial z \partial v}\
\frac{\partial^2 \bar {\cal F}_1(\bar \zeta)}{\partial \bar w \partial \bar u}
= \frac{G^{ij}  C_{J \bar J M_i} C_{J \bar J M_j}}{(z-v)^2 (\bar w - \bar u)^2}\,.
\label{extractingM}
\fe  
Therefore, we can isolate the contributions from marginal operators on the RHS of  \eqref{J4pfs} as
\ie 
&\langle   \bar J (  w) J(z) \bar J(u)
 J(  v)\rangle  
= \frac{G^{ij}  C_{J \bar J M_i} C_{J \bar J M_j}}{(z-v)^2 (\bar w - \bar u)^2}
+
\sum_{\phi\neq 1, M_i}
\A_\phi^2 \
\frac{\partial^2 {\cal F}_{h_\phi}(\zeta)}{\partial z \partial v}\
\frac{\partial^2 \bar {\cal F}_{\bar h_\phi}(\bar \zeta)}{\partial \bar w \partial \bar u}+ O(\Delta)\, ,
 \label{schannel}
\fe 
where we have defined,
\ie 
\A_\phi^2\equiv  \lim_{\Delta \to 0}\frac{|C_{{\cal O}{\cal O} \phi}|^2}{\Delta^2}\,.
\label{alphaphidef}
\fe 
In the following, we will show that for operator $\phi$ that does not asymptote to the identity or marginal operators,
$\A_\phi$ must vanish at least as $\Delta^{1/2}$, as a consequence of the consistency of the four-point function in the limit $\Delta \to 0$.

In the $t$-channel, the same four-point function $\langle \bar J (  w)  J(z)\bar J(u)  J(  v)\rangle$ can be expanded as
\ie
\langle   \bar J (  w) J(z) \bar J (u)
 J ( v)\rangle
= \frac{1}{(z-v)^2 (\bar w - \bar u)^2}+
\sum_{\phi\neq 1}\A_\phi^2\
\frac{\partial^2 {\cal F}_{h_\phi}(1-\zeta)}{\partial z \partial v}\
\frac{\partial^2 \bar {\cal F}_{\bar h_\phi}(1-\bar \zeta)}{\partial \bar w \partial \bar u}
+ O(\Delta)\,,
\label{J4pft}
\fe 
where we have isolated the identity contribution in the first term on the RHS using \eqref{OtoJJb}.  By combining the identity, 
\begin{equation}
(z-v)^2 \frac{\partial^2 f(\zeta)}{\partial z \partial v} = - (1-\zeta)^2 \frac{d}{d \zeta} \left[
\zeta \,\frac{d f(\zeta)}{d \zeta} \right]\,,
\end{equation}
which holds for any function $f(\zeta)$ of $\zeta$,
with the defining differential equation for hypergeometric functions,
\begin{equation}
\left[ \zeta (1-\zeta) \frac{d^2}{d\zeta^2}
+(2h - (2h+1) \zeta) \frac{d}{d\zeta} - h^2\right] {}_2 F_1(h, h, 2h; \zeta)=0\,,
\label{hyperdefeqn}
\end{equation}
we find that the $t$-channel global conformal blocks 
${\cal F}_{h}(1-\zeta)$ from \eqref{globalblock} are eigenfunctions of 
$(z-v)^2 \partial_z \partial_v$ as below,
\begin{equation}
 (z-v)^2  
 \frac{\partial^2 {\cal F}_{h}(1-\zeta)}{\partial z \partial v}= - (1-\zeta)^2 \frac{d}{d \zeta} \left[
\zeta \frac{d }{d \zeta} {\cal F}_{h}(1-\zeta) \right]
= h (1 - h) {\cal F}_{h}(1-\zeta)\,.
\label{eigenhypergeo}
\end{equation}
Thus, we can write the $t$-channel expansion (\ref{J4pft}) as,
\begin{equation}
\langle   \bar J (  w) J(z) \bar J (u)
 J ( v)\rangle
= \frac{1 + \sum_{\phi\neq 1}\A_\phi^2\ h_\phi\bar h_\phi  (1 - h_\phi)
(1 - \bar h_\phi)
{\cal F}_{h_\phi}(1-\zeta)\bar {\cal F}_{\bar h_\phi}(1-\bar \zeta)}{(z-v)^2 (\bar w - \bar u)^2} +O(\Delta)\,.
\label{holomorphicfactorization}
\end{equation}
As a consequence of \eqref{OtoJJb}, 
$J$ and $\bar J$ are holomorphic and anti-holomorphic respectively to the leading order in $\Delta$. By applying the differential operators $(\bar z-\bar v)^2 \bar\partial_{\bar z} \bar \partial_{\bar v}$ and $(u-w)^2  \partial_{u} \bar \partial_{w}$ to the above equation and using the analogs of the eigenvalue equation \eqref{eigenhypergeo}, we find 
\ie 
\sum_{\phi\neq 1}\A_\phi^2 \ h_\phi^2\bar h_\phi^2  (1 - h_\phi)^2
(1 - \bar h_\phi)^2
{\cal F}_{h_\phi}(1-\zeta)\bar {\cal F}_{\bar h_\phi}(1-\bar \zeta)=O(\Delta)\,.
\label{holofact}
\fe
It then follows from the convergence of the $t$-channel OPE \cite{Pappadopulo:2012jk,Hogervorst:2013sma,Qiao:2017lkv} and the positivity of the conformal blocks ${\cal F}_{h_\phi}(1-\zeta)\bar {\cal F}_{\bar h_\phi}(1-\bar \zeta)$ for $\zeta=\bar\zeta \in \mR_+$ \cite{Fitzpatrick:2012yx} that $\alpha_\phi$ must vanish in the limit
unless the conformal weights of $\phi$ obey $h_\phi\bar h_\phi  (h_\phi-1)
(\bar h_\phi-1)=0$.

Since the spin of $\phi$ is quantized,
$\alpha_\phi$ can be non-zero only if 
\begin{equation}
    (h_\phi, \bar h_\phi) \in \{ (n, m) \in \mZ_{\geq 0} \otimes \mZ_{\geq 0}\,|\,
    nm(n-1)(m-1)=0 \} \ .
\end{equation}
With this restriction and the fact that ${\cal F}_{0}(\zeta)=1$, only conformal families with primary conformal weights $(n,1)$ and $(1,n)$ with $n \geq 2$ contribute in the $s$-channel expansion \eqref{schannel} and consequently,
\ie
&\langle   \bar J (  w) J(z) \bar J(u)
 J(  v)\rangle  \\
&= \frac{G^{ij}  C_{J \bar J M_i} C_{J \bar J M_j}}{(z-v)^2 (\bar w - \bar u)^2}
+
\frac{ \sum_{n=2}^\infty \beta_n \partial_z \partial_v{\cal F}_{n}(\zeta) }{(\bar w - \bar u)^2}
+
\frac{ \sum_{n=2}^\infty \bar \beta_n \partial_{\bar w} \partial_{\bar u}\bar {\cal F}_{n}(\bar \zeta) }{(z-v)^2}
+ O(\Delta)\, ,
\fe
where 
\begin{equation}
\beta_n = \sum_{\phi: (h_\phi, \bar h_\phi) = (n, 1)} \alpha_\phi^2\,, ~~{\rm and}~~
\bar \beta_n = \sum_{\phi: (h_\phi, \bar h_\phi) = (1, n)} \bar \alpha_\phi^2 \,.
\end{equation}
On the other hand, the $t$-channel expansion gives simply 
$\langle   \bar J (  w) J(z) \bar J(u)
 J(  v)\rangle = (z-v)^{-2} (\bar w - \bar u)^{-2} + O(\Delta)$ as we have seen in the above.\footnote{\label{footnote:correction}The fact that the correction is of order $O(\Delta)$ as opposed to $O(\Delta^{1/2})$ follows from \eqref{OtoJJb} and the assumption that the correlators among $J,\bar J,K$ are well-defined in the limit $\Delta \to 0$. In particular the would-be $O(\Delta^{1/2})$ contribution to the four-point function $\langle   \bar J  J \bar J J\rangle$ is proportional to  $\langle   K  J \bar J J\rangle$ or $\langle    \bar J K \bar J J\rangle$ (up to permutations) which vanish in this limit. See also \eqref{holofact}.} The crossing invariance of the four-point function therefore demands,
\ie
&     G^{ij}  C_{J \bar J M_i} C_{J \bar J M_j-1}-1\\
 &    = (z-v)^2 \sum_{h_\phi=2}^\infty \alpha_\phi^2\partial_z \partial_v{\cal F}_{h_\phi}(\zeta)
     + (\bar w - \bar u)^2 \sum_{\bar h_\phi=2}^\infty \alpha_\phi^2\partial_{\bar w} \partial_{\bar u}\bar {\cal F}_{\bar h_\phi}(\bar \zeta) +O(\Delta) \\
& = 
(1-\zeta)^2 \frac{d}{d \zeta} \left[
\zeta \ \frac{d}{d \zeta}  \sum_{n=2}^\infty \beta_n
 {\cal F}_{n}(\zeta) \right]
 + (1-\bar \zeta)^2 \frac{d}{d \bar \zeta} \left[
\bar\zeta \ \frac{d}{d \bar\zeta}  \sum_{n=2}^\infty\bar \beta_n
 \bar{\cal F}_{n}(\bar\zeta) \right] +O(\Delta) \,.
 \label{comparingsandt}
     \fe

To solve the above bootstrap equation at leading order in $\Delta$, we define the following linear functional by a contour integral around $\zeta=0$,
\ie 
\omega_n (f(\zeta))=\oint_{\zeta=0} {d\zeta\over 2\pi i } {1\over n(n-1)\zeta(1-\zeta)}\left (\cF_{1-n}(\zeta)+{{d\over d\zeta}\cF_{1-n}(\zeta)\over n(n-1)}  \right)  f(\zeta)\,.
\label{functional}
\fe
It implements the projection to the contribution from the conformal families with primary conformal weight $(n,2)$ in the $s$-channel in \eqref{comparingsandt} thanks to the following property
\ie 
\omega_n\left ( (1-\zeta)^2 {d\over d\zeta} \left[\zeta {d\over d\zeta} \cF_m(\zeta)  \right] \right) = \D_{m,n}\,.
\fe
This follows from basic hypergeometric identities such as \eqref{eigenhypergeo} (see \cite{Heemskerk:2009pn} for a similar projection functional). A similar linear functional can be defined by the corresponding contour integral in $\bar\zeta$. 

As explained in \cite{Pappadopulo:2012jk,Hogervorst:2013sma,Qiao:2017lkv}, the $s$-channel conformal block expansions in unitary theories, such as that in \eqref{comparingsandt}, are uniformly convergent for $\zeta,\bar\zeta$ treated as separate complex variables valued in the cut-plane $\mC\backslash  [1,\infty)$. Furthermore, since $\cF_n$ and $\bar \cF_n$  are holomorphic in the cut-plane, the sum in \eqref{comparingsandt} converges to holomorphic functions of $\zeta$ and $\bar\zeta$ respectively. 
We can therefore apply the linear functional defined in \eqref{functional} to both sides of \eqref{comparingsandt}.\footnote{Using the fact that the expansions in \eqref{comparingsandt} come with non-negative coefficients $i.e.$ $\beta_n,\bar \beta_n\geq 0$, we can exchange the integral and the sum by the dominated convergence theorem.}
Consequently we conclude $\beta_n=\bar \beta_n=0$ with $n\geq 2$ to this order, which implies 
\ie
\alpha_\phi^2=O(\Delta)\,,
\label{alphaphismall}
\fe
for any operator $\phi$ that does not approach either identity or marginal operators in the limit, and 
\ie
     &G^{ij}  C_{J \bar J M_i} C_{J \bar J M_j} = 1 +  O(\Delta)\,.
     \label{normalization}
\fe

Since $G_{ij}$ is positive definite, \eqref{normalization} 
implies that 
$|C_{J \bar J M_i}| $ are bounded above  for all $i$ in the limit. In particular,  
since $M$ is a normalized hermitian linear combination of the exactly marginal operators $M_a$,
we conclude that 
\ie 
 C_{J \bar J M} \leq  1 +  O(\Delta)\,,
    \label{cjjbarM}
    \fe 
which is what we wanted to show.\footnote{Note that we do not assume the existence of a local stress tensor in the proof. Therefore, the results in this section also apply to the conformal manifolds of surface defects in $d\geq 3$ CFTs.}
As a by-product, we also deduce from \eqref{alphaphidef} and \eqref{alphaphismall} that, the OPE coefficient between the operator $\phi$ and the light operator $\cO$ satisfies
\begin{equation}
    |C_{{\cal O}{\cal O}\phi}| = O(\Delta^{3/2}) \,,
\end{equation}
except for $\phi$ that asymptotes to the identity or marginal operators in the limit of vanishing gap.

The inequality  \eqref{cjjbarM} not only establishes that vanishing gap requires infinite distance on the conformal manifold, but also constrains the rate at which the conformal dimension gap $\Delta$ approaches zero as follows.
Using Riemann normal coordinates $t^a$ dual to $M_a$ near the limit of vanishing gap and introducing $\alpha_a$ as the limit of $C_{J \bar J M_a}$, 
equations (\ref{ST}) and (\ref{JbarJMOPE}) can be generalized to
\begin{equation}
    \frac{\partial }{\partial t^a}
    \log \Delta = - C_{J \bar J M_a}
    (1 + O(\Delta))
    = - \alpha_a (1 + O(\Delta))\,.
\end{equation}
We can then integrate this to obtain
\begin{equation}
\Delta =
\exp( -\alpha_a t^a + O(1) )\,.
\label{multimodulione}
\end{equation}
By equation (\ref{normalization}), 
the length $||\alpha||\equiv \sqrt{G^{ab}\alpha_a\alpha_b}$ of the vector $\alpha_a$ is bounded
above, $||\alpha||\leq 1$, and the bound is saturated if and only if $C_{J \bar J M_i}=0$ for all non-exactly-marginal directions ($i.e.$ $i>\dim \cM$). 

It is convenient to parametrize the Riemann normal coordinates $t^a$ as $t^a = e^a t$, where $t$ is the geodesic distance and $e^a$ is a unit vector defined by
\begin{equation}
    e^a = \cos \theta \ G^{ab} \alpha_b + \sin \theta \
    e_\perp^a\,,
\end{equation}
where $e_\perp^a$ is a unit vector satisfying $\alpha_a e_\perp^a=0$. We choose $0 \leq \theta \leq \pi/2$ so that the geodesic length grows toward the direction of $G^{ab}\alpha_b$. With this parametrization, equation (\ref{multimodulione}) becomes 
\begin{equation}
\Delta = \exp(- \alpha t + O(1))\,,
\end{equation}
with $\alpha = \cos \theta||\alpha|| \leq 1$. The upper bound is saturated if and only if  $\theta = 0$ and $||\alpha||=1$, 
when 
the geodesic points in the direction of $G^{ab} \alpha_b$ and
 all non-exactly-marginal directions have $C_{J\bar JM_i}=0$.\footnote{In other words, the direction $G^{ab} \alpha_b$ is the fastest way to achieve vanishing gap.}

\section{Properties of Limiting CFT}
\label{sec:limitingCFT}

Here we describe features of the CFT in the limit of vanishing gap $\Delta_{\rm gap}\to 0$ as one moves on its conformal manifold $\cM$. In particular we will derive the emergence of a large target space in this limit, establishing Theorem~\ref{thm:CFTinthelimit} and~\ref{thm:lowerbound}.

In the previous section, we have focused our attention to one of the primary operators with vanishing conformal dimensions. In general there exists a set of
linearly independent primary operators $\{ {\cal O}_n \}$ whose conformal dimensions vanish
simultaneously. The CFT may also contain other primary operators whose conformal dimensions remain finite or diverge in this limit. It is convenient to set a scale $\Delta_{\rm finite}$ such that the conformal dimensions of all such states are greater than $\Delta_{\rm finite}$ in the limit.\footnote{In particular, $\Delta_{\rm finite}\leq 1$ since the descendant of an operator $\cO$ with vanishing dimension may become a primary operator in the limit as we saw in Section~\ref{sec:zerogaptoinfdist}.}

For each such primary operator ${\cal O}_n$, we define its normalized descendant $J_n$ by
\begin{equation}
    \partial {\cal O}_n = i\sqrt{\Delta_n} J_n\,,
    \label{Jn}
\end{equation}
where ${\cal O}_n$ and $J_n$
have conformal weights $(\frac{1}{2} \Delta_n, \frac{1}{2} \Delta_n)$
and $(\frac{1}{2} \Delta_n +1, \frac{1}{2} \Delta_n)$  respectively, and similarly for $\bar J_n$ and $K_n$ as in \eqref{OtoJJb}. As we explain below, the operators $J_n,\bar J_n$ are emergent currents in the limit of vanishing gap while $K_n$ are emergent marginal operators, but they are not necessarily all linearly independent in the limit. 
We will focus on the operators $J_n$ and the linear relations among them that emerge in the limit $\Delta_n \to 0$.  

For example, the $S^1$ sigma-model described by a periodic scalar field $X\sim X + 2 \pi R$
contains the infinite tower of
momentum eigenstates  whose conformal dimensions vanish in the large radius limit  
$R \rightarrow \infty$. In this example,
it is convenient to use a complex basis for these light operators,
\begin{equation}
    {\cal O}_n=\exp\left( i \frac{n}{R} X\right)\,,\quad n\in \mZ\,,
\end{equation}
with scaling dimension $\Delta_n=(n/ R)^2$ and correspondingly a complex basis for the descendants,
\begin{equation}
\partial {\cal O}_n = \sqrt{\Delta_n} J_n\,,
~~~
\partial {\cal O}_{-n} = -\sqrt{\Delta_{-n}} J_{-n}\,,
 \quad n\in \mZ_+\,,
    \label{S1Jn}
\end{equation}
with
\begin{equation}
    J_{n} =  i \partial X
    \exp\left(   i \frac{n}{R} X\right)\,,
  \quad  n \in \mZ \,.
\end{equation}
Though the operators $J_n$ are linearly independent at finite $R$,
they all become $ i\partial X$  in the limit of $R \rightarrow \infty$. 

To understand in general the emergent linear relations among $J_n$'s in the limit $\Delta_n \to 0$, it is useful to consider the operator product algebra of ${\cal O}_n$'s,
\begin{equation}
    {\cal O}_n(z) {\cal O}_m(w) =\sum_k
    C_{nm}^k |z-w|^{\Delta_k - \Delta_n - \Delta_m} {\cal O}_k(w)
    + O(|z-w|^{\Delta_{\rm finite}}).
    \label{OPE}
\end{equation}
Here we are assuming that the operators ${\cal O}_n$ are hermitian. The OPE is dominated by light operators that have vanishing scaling dimensions in the limit (the first term on the RHS of \eqref{OPE}). 
To derive linear relations among $J_n$'s, we act
by $(\partial_z + \partial_w)$ on both sides  of \eqref{OPE}, 
taking the limit $\Delta_n \rightarrow 0$, 
and then set $ z \to w$. Since we assume the light operators to have well-defined sphere correlation functions in the limit, 
$C_{nm}^k$ cannot diverge and we obtain the following linear relations,
\begin{equation}
\sqrt{\Delta_n}  J_n (z) +  \sqrt{\Delta_m} J_m(z) = 
\sum_k C_{nm}^k \sqrt{\Delta_k}
 J_k(z)\,.
\label{relations}
\end{equation}

For example, in the $S^1$ sigma-model example discussed above, these relations for
$J_n$ defined as in \eqref{S1Jn} become
\begin{equation}
n J_n + m J_m = 
    (n+m) J_{n+m}\,,
\end{equation}
and the unique solution is 
$J_{n} = J_1$.
For the $S^1$ sigma-model, \eqref{relations} clearly give a complete set of linear relations among $J_n$'s in the $\Delta_n\to 0$ limit. Since the operator product algebra \eqref{OPE} contains all the information about operatorial relations among ${\cal O}_n$'s, we
expect that this is the case in general, $i.e.$, all the linear relations among $J_n$'s defined by \eqref{Jn} are given by \eqref{relations}.

We can also derive the operator product algebra of the operators $J_n$ by 
acting 
$\partial_z \partial_w$ on both sides of equation (\ref{OPE}) and taking the limit $\Delta_n\to 0$ and then $\bar w \rightarrow \bar z$ to suppress the $O(|z-w|^{\Delta_{\rm finite}})$ terms. We obtain the following,
\begin{equation}
\label{JJOPE}
    J_n(z) J_m(w) = \sum_k C_{nm}^k
    \frac{\Delta_k-\Delta_n- \Delta_m}{\sqrt{\Delta_n\Delta_m}}
    \left( 
    \frac{1}{(z-w)^2} -\frac{i\sqrt{\Delta_k}}{z-w} J_k(w)
    \right) + {\rm regular}\,.
\end{equation}
The unconventional coefficient of $1\over (z-w)^2$ is due to
the linear relations \eqref{relations} among $J_n$'s. 
By choosing an orthonormal hermitian basis,
\begin{equation}
    {\cal J}^\mu = \sum_k U^{\mu n} J_n\,,
    \label{linearcomb}
\end{equation}
taking into account the relations  \eqref{relations}, we can 
put their OPE in the standard form below,
\begin{equation}
{\cal J}^\mu (z)
    {\cal J}^\nu(w) = \frac{\delta^{\mu\nu}}{(z-w)^2} +  \frac{if^{\mu\nu}_\rho}{z-w}{\cal J}^\rho(w) 
    + {\rm regular}\,,
    \label{currentalgebra}
\end{equation}
for some constant $f^{\mu\nu}_\rho$, valid in the limit $\Delta_n \to 0$. In the following, we take the number of linearly independent emergent currents to be $N$.

Note that $f^{\mu\nu}_\rho$ would vanish in the limit if 
 the ratio $\Delta_n/\Delta_m$ for any pair $n, m$ remains non-zero and finite, due to the additional factor of $\sqrt{\Delta_k}$ in the $(z-w)^{-1}$ term on the RHS of \eqref{JJOPE}.
Although there is a logical possibility that
different $\Delta_n$ are scaled in differently so that a nonzero $f^{\mu\nu}_\rho$ survives in the limit, we will argue below that unitarity requires $f^{\mu\nu}_\rho$ to vanish identically. 

The OPE \eqref{currentalgebra} defines a Kac-Moody algebra.
Since the Killing form  for the algebra is $\delta^{\mu\nu}$ and positive definite,  consistency demands that $f^{\mu\nu}_\rho$ is proportional to the structure constant of a compact Lie group $G$. The conventional Kac-Moody currents obey a normalization such that the leading OPE singularity is ${k\over 2}(z-w)^{-2}$  where $k\in \mZ_+$ is the level of the Kac-Moody algebra. Comparison to \eqref{currentalgebra} indicates that
$f^{\mu\nu}_\rho$ is given by the structure constant of $G$ multiplied by the factor $\sqrt{2\over k}$.

Let us now show that $f^{\mu\nu}_\rho=0$, which amounts to saying that the underlying group $G$ is abelian.
Suppose to the contrary that the group $G$ contains a simple non-Abelian subgroup, for which
$f^{\mu\nu}_\rho$ does not vanish. It is well-known that the Kac-Moody algebra generators can be constructed using massless free scalar fields and parafermion fields \cite{Gepner:1987sm}.
In particular, the current ${\cal J}_H$ in a direction $H$ of the maximum torus of the group $G$ normalized as in \eqref{currentalgebra} is given by,
\begin{equation}
    {\cal J}_H = i \partial X\,,
\end{equation}
for a periodic scalar field $X$, which in the limit becomes free with the two-point function $\partial X(z) \partial X(w) \sim -(z-w)^{-2}$. Because of \eqref{Jn} and \eqref{linearcomb}, there must be a light primary operator ${\cal O}$ of scaling dimension $\Delta_\cO$
such that 
\begin{equation}
    \partial {\cal O} =q  {\cal J}_H = i q \partial X\,,
\end{equation}
for some real number $q$ near the limit  $\Delta_\cO\to 0$. By integrating this equation, we obtain
\begin{equation}
     {\cal O}(z, \bar z) = e^{iq X} {\cal S}(\bar z)\,,
\end{equation}
for some operator ${\cal S}(\bar z)$ that only depends on $\bar z$. Since the scaling dimension of such an operator is bounded from below $\Delta_\cO\geq q^2$, we need $q$ to vanish in the limit. This is a contradiction since $q$ is the charge carried by ${\cal O}$ with respect to the current ${\cal J}_H$ and 
must be quantized in any unitary representation of a simple compact non-Abelian Lie group. Therefore, the group $G$ cannot contain any non-Abelian subgroup (with finite Kac-Moody level $k$), and $f^{\mu\nu}_\rho$ must vanish as claimed earlier.

Since $f^{\mu\nu}_\rho=0$,
all the currents that descend from the light operators in the limit can be represented as
\begin{equation}
{\cal J}^\mu (z)= i\partial X^\mu\,,
\label{normJ}
\end{equation}
in terms of free bosons $X^\mu$. 
For each $X^\mu$, there must be a primary operator ${\cal O}$ such that
\begin{equation}
    \partial {\cal O} = i q \partial X^\mu\,,
    \label{OtoX}
\end{equation}
for some real number $q$. 
Since the scaling dimension of ${\cal O}$
vanishes in the limit, we must also have
\begin{equation}
    \bar \partial {\cal O} = i \bar q \bar \partial \bar X^\mu\,,
\end{equation}
for another free boson $\bar X^\mu$ and another real number $\bar q$. 
By integrating the above two equations, we obtain the following representation of the light operator $\cO$,
\begin{equation}
    {\cal O} = e^{i q X_L^\mu (z) + i \bar q \bar X_R^\mu(\bar z)}\,,
\end{equation}
where $X_L^\mu$ is a projection of $X^\mu$ to its holomorphic part and
$\bar X_R^\mu$ is a projection of $\bar X^\mu$ to its anti-holomorphic part. 
The fact that ${\cal O}$ has to be a scalar requires $\bar q=\pm q$. Since $q \rightarrow 0$ in the limit, mutual locality of operator spectrum requires that only one of the two possibilities is realized which we take to be $q=\bar q$. Consequently we can write,
\begin{equation}
    {\cal O} = e^{i q X^\mu}\,,
    \label{O2rep}
\end{equation}
where $X^\mu \equiv  X_L^\mu + \bar X_R^\mu$. 
By taking OPE of ${\cal O}$ above with itself, we can generate an infinite tower of light operators $ e^{i n q X^\mu}$ 
for any positive integer $n$, and  $e^{-i n q X^\mu}$ from their hermitian conjugates. Since we can approximate any real number $p$ by $nq$ with $q\to 0$ and $n\to \pm \infty$, 
we find that the spectrum of the theory must contain $e^{i p X^\mu}$ for any real number $p$ in the limit. One can identify $p$ as the momentum charge for the current $\cJ^\m$ in \eqref{OtoX} in the limit.

By repeating the above procedure for all currents ${\cal J}^\mu$ that emerge in the limit and taking products of the resulting primary momentum operators, we find that the CFT in the limit contain
primary operators of the form
\begin{equation}
    {\cal O}_{p_\m}(z, \bar z) = e^{i \sum_\mu p_\mu X^\mu (z, \bar z)}\,,
    \label{momentumops}
\end{equation}
for any real $N$-vector $p_\mu$. They describe a continuum of scaling dimensions $\Delta = \sum_\mu p_\mu^2$ without a gap above the vacuum. It shows that the 
 CFT in this limit contains a subsector of local operators described by the sigma-model with a non-compact target space $\mR^N$. 

The limiting CFT may contain other primary operators with scaling dimensions at or above $\Delta_{\rm finite}$. Moreover,  
the limit may not necessarily factorize into a tensor product of
the $\mR^N$ sigma-model and a compact CFT. Instead they can be coupled together by an orbifold or a more general fibration.

As an example, let us consider the orbifold CFT on $S^1/\mZ_2$, where the compact boson $X$ of radius $R$ is subjected to an identification by the $\mZ_2$ reflection $X \sim - X$. The limit of vanishing gap here is obviously $R\to \infty$ (up to choosing a duality frame) and the limiting theory is described by the $\mZ_2$ orbifold of the $\mR$ sigma-model. In this case, the emergent current from \eqref{Jn} lives in the twisted sector of a dual symmetry which we explain below.
In the untwisted sector, the conformal dimension $\Delta_n =  ( n/ R )^2$ of the $\mZ_2$-invariant operator
${\cal O}_n=\sqrt{2}\cos(n X/R)$ with $n\in \mZ_+$ vanishes in the large radius limit $R \rightarrow \infty$. The operator $J_n$ as defined in \eqref{Jn} is given by
\ie
J_n = i  \partial X \widetilde{\cal O}_n\,,
\fe
where $\widetilde{\cal O}_n=\sqrt{2} \sin(n X/ R)$.\footnote{It is properly normalized as $\widetilde{\cal O}_n(z)\widetilde{\cal O}_n(w) \sim |z-w|^{-2\Delta_n} - {1\over \sqrt{2}}|z-w|^{2\Delta_n}{\cal O}_{2n}(w) + \cdots$.} Though $\widetilde{\cal O}_n$ may seem to vanish in the $R\rightarrow \infty$ limit, it remains a nontrivial operator in the limit.
Since it has a nontrivial monodromy relation with operators in the $\mZ_2$ twisted sector, the proper way to think about it is a topological ($i.e.$ dimension 0) operator at
the end-point of the topological defect line that implements the quantum $\widehat\mZ_2$ symmetry of the orbifold \cite{Vafa:1989ih}. Another way to say this is that, the operator $i \partial X$ belongs to the twisted sector of the quantum $\widehat\mZ_2$ symmetry and thus is attached to the  $\widehat\mZ_2$ topological line (which ends on $\widetilde \cO_n$). 
More generally, the emergent currents 
\eqref{normJ} should be interpreted as descending from operators in the twisted sector of a topological defect line $\cL$ and consistency requires the ground state in the $\cL$-twisted sector to have zero conformal dimension in the limit $\Delta \to 0$.\footnote{A necessary condition for this to happen is to have a trivial spin selection rule in the $\cL$-twisted sector, which is equivalent to the anomaly free condition for invertible symmetries \cite{Chang:2018iay}.} 

In the example above, the compact sector in the limit is trivial (simply a gapped vacuum). As another simple example but with a nontrivial compact sector, we can consider the $SU(2)_k$ CFT with the exactly marginal deformation $M=J_3\bar J_3$. Using the orbifold equivalence
\ie 
SU(2)_k \cong {SU(2)_k/U(1)\times U(1)_{2k}\over \mZ_k}\,,
\label{WZWexample}
\fe
between the  $SU(2)_k$ CFT and a product of the $\mZ_k$ parafermion CFT and the compact boson CFT at radius $R=\sqrt{2k}$ where $\mZ_k$ acts diagonally on the two factors,\footnote{Here the $\mZ_k$ symmetry of the $SU(2)_k/U(1)$ coset CFT  descends from the standard $\mZ_k$ symmetry of the parafermions 
and that of the $U(1)_{2k}$ CFT is the non-anomalous $\mZ_k$ momentum (shift) symmetry.} we can identify this one-dimensional conformal manifold generated by $M$ as that for the compact boson factor. Consequently, in the limit of vanishing gap which corresponds to large radius limit of the compact boson, the limiting CFT is described by the $\mZ_k$ parafermion CFT and the $\mR$ sigma-model coupled together by the $\mZ_k$ orbifold.\footnote{More precisely, there are $k$ copies of the $\mR$ sigma-model related by the $\mZ_k$ symmetry that is gauged.}

After taking into account the above caveat, what we have shown in this section is that 
the limiting CFT contains the operator algebra of $N$ free non-compact bosons as a subalgebra of its local operators, modulo such topological defect lines. 
We conclude that
the full CFT central charge $c$ puts the upper bound 
$N \leq c$ on the number $N$ of emergent non-compact directions in the limit of vanishing gap. We have thus established Theorem~\ref{thm:CFTinthelimit} regarding the behavior of the CFT in this limit. 

Let us move on to prove Theorem~\ref{thm:lowerbound}
about the lower bound on $\alpha$.
Marginal operators in the limit consists of either 
\begin{equation}
     M^{\mu\nu}(z, \bar z) = \partial X^\mu \bar{\partial} X^\nu \,,
     \label{marginalatlimit}
\end{equation}
which make a linearly independent basis of the marginal operators 
emergent  from $\partial \bar \partial {\cal O}_{p_\mu}$ by \eqref{OtoJJb} and \eqref{momentumops}
or operators which entirely commute with $\partial X^\mu$.
For $M^{\mu\nu}$, the three-point function with the emergent currents (see \eqref{normJ})
$\cJ^\rho$ and $\bar \cJ^\sigma$ is
\begin{equation}
    \langle \cJ^\rho(z) \bar \cJ^\sigma(\bar w) M^{\mu\nu}(u, \bar u)\rangle= 
    \frac{\delta^{\rho\mu}\delta^{\sigma \nu}}{(z-u)^2(\bar w - \bar u)^2}\,,
    \label{freethreepoint}
\end{equation}
up to terms that vanish in the $\Delta \to 0$ limit.
This is consistent with 
\eqref{cjjbarM} as expected.
The marginal operators which commute with $\partial X^\m$ have zero three-point functions with ${\cal J}^\rho$ and do not contribute to our estimate of $\alpha$.
Note that these operators $M^{\m\n}$ do not in general survive as exactly marginal operators away from the limit of vanishing gap. In the example of quintic SCFT in the large volume limit discussed in Section~\ref{sec:quinticexample}, the number of decompactifying directions is $N=6$ but only 1 out of the 36 operators is exactly marginal, which corresponds to the overall volume deformation in the limit.

The exactly marginal operator $M$ that couples to $t$ in 
Theorem~\ref{thm:gapvanishing} should become a linear
combination of $M^{\mu\nu}$ in the limit,
 \ie 
M=\sum_{\m, \n } \kappa_{\mu\nu} M^{\mu\nu}
\, .
 \fe
Since $M$ is canonically normalized, 
the coefficients $\kappa_{\mu\nu}$ are normalized as
\ie
\sum_{\m, \n } (\kappa_{\mu\nu})^2 = 1 \, .
\label{kappanormalization}
\fe
For an operator described by  \eqref{momentumops} with momentum $p$ in the limit, the decay coefficient $\A(p)$ for its conformal dimension $\Delta = \exp(- \alpha(p) t + O(1))$ in the direction of $M$ is given by
\ie
\alpha(p) = \frac{\sum_{\m\n} \kappa_{\m\n} p_\m p_\n}{\sum_\m (p_\m)^2} \,,
\label{alphap}
\fe
which follows from \eqref{ST}. Below we will derive a lower bound on the maximum decay coefficient $\A(p)$ that can be achieved for any $p$. Intuitively this amounts to picking the operator whose conformal dimension decays the fastest along the  geodesic given by $M$, and consequently the lower bound is an intrinsic property of $M$  that specifies the approach to the limit.

The sector of the CFT described by the $N$ noncompact free bosons has the standard positive-definite kinetic terms 
and is parity invariant. Since we add
 ${t\over 2\pi}\int d^2 x M(x)$ to the CFT action and take $t\rightarrow \infty$ to reach the limit, the integral of $M$ should be a positive and parity preserving operator. 
 In general, $M$ itself may have a parity-odd component (from the antisymmetric part of $\kappa_{\m\n}$) but that is a total derivative in the limit. Here we assume that in a neighborhood of the limit, we can choose $M$ such that it is parity even.\footnote{Note that this does not require the whole CFT to be parity invariant (see the example discussed around \eqref{WZWexample}).}
 Correspondingly $\kappa_{\mu \nu}$ should be symmetric and positive semi-definite.
 
 It then follows from standard linear algebra that $\alpha(p)$ in \eqref{alphap} is bounded from above by the largest eigenvalue of $\kappa_{\m\n}$ and the bound is saturated when $p$ is aligned with the corresponding eigenvector. Since $\kappa_{\m\n}$ is normalized by \eqref{kappanormalization}, its largest eigenvalue is bounded from below by $1/ \sqrt{N}$. Therefore we conclude that 
 \ie 
{1\over \sqrt{N}} \leq \max{}_{p \in \mR^N}\A(p)\,.
 \fe
Combining with the unitarity bound $N \leq c$, 
we have shown that, given an exactly marginal operator $M$ that specifies the limit to vanishing gap, there is always an operator 
for which the decay rate 
$\alpha$ of its conformal dimension is bounded below by
 \ie 
 {1\over \sqrt{c}}\leq \alpha \,.
\label{alphalowerbound}
 \fe
The bound is saturated if and only if $c=N$ 
and $\kappa_{\m\n} = \D_{\m\n}/\sqrt{N}$ so that 
the exactly marginal operator $M$ scales  uniformly 
the $N$ emergent decompactifying directions.\footnote{The inequality can be improved when $c$ is fractional, in which case $\lfloor 
 c\rfloor\geq N$ and $\lfloor 
 c\rfloor^{-1/2} \leq \A$. When $c\in \mathbb{Q}$, this improved inequality can be saturated only if $c-N={6\over (m+2)(m+3)}$ for $m\in \mZ_+$ so that the compact sector is described by a Virasoro minimal model. When $c$ is irrational, this inequality is always strict.
 } 

 The above analysis generalizes straightforwardly if extra conserved currents are preserved as the CFT approaches the $\Delta_{\rm gap} \to 0$ limit. For example, if the limit is supersymmetric,
 the emergent scalar fields $X^\m$ must be paired with real fermions $\psi^\m$. It is conventional to write the superconformal CFT (SCFT) central charge as $c=3\hat c/2$. Then the compact sector of the limiting CFT is also an SCFT with central charge $\hat c_{\rm rest} =\hat c- N$ and consequently the lower bound \eqref{alphalowerbound} on $\A$ is strengthened to $\hat c^{-1/2} \leq \A$.

As discussed in the Introduction, this lower bound can be translated in the AdS units as
\begin{equation}
    \left(L_{\rm Planck}\right)^{1/2}
 \leq \ \alpha_{\rm AdS}\,.
\end{equation}
As explained in Section~\ref{sec:gravityimplications}, after taking a flat space limit, this coincides with the Sharpened Distance Conjecture \eqref{SDC} for $D=6$ \cite{Etheredge:2022opl}. 
For example, let us consider the supersymmetric sigma-model on ${\rm Sym}^N(T^4)$, which
is dual to Type IIB string theory on 
${\rm AdS}_3 \times {\rm S}^3 \times T^4$.
In this case, 
the lower bound for $\alpha_{\rm AdS}$ can be saturated, for example, by Kaluza-Klein modes in the large volume limit of $T^4$ (see Section~\ref{sec:gravityimplications}).

On the other hand, consider the bosonic 
sigma-model on ${\rm Sym}^N (T^4)$. This model is conformal 
at least on the orbifold locus (the orbifold blowup modes are not exactly marginal). Without supersymmetry,
the lower bound on $\alpha$ is weaker as in \eqref{alphalowerbound}, which in the AdS units is 
\begin{equation}
    \left(\frac{2}{3}L_{\rm Planck}\right)^{1/2} \leq \alpha_{\rm AdS} .
\end{equation}
This lower bound is saturated in the large volume limit of the seed $T^4$ sigma-model. Though the Sharpened Distance Conjecture is violated in this example,
we do not claim it as a strict counter-example since  it is not clear
whether
the large $N$ limit of this bosonic sigma-model 
has a weakly coupled gravitational dual in ${\rm AdS}_3$.

The analysis in this section shows that, once there is one primary operator with $\Delta \rightarrow 0$, there appears 
an infinite tower of operators whose conformal weights vanish simultaneously.
In the bulk AdS, there is a corresponding tower of light states. 
Since $\alpha_{\rm AdS}$ is bounded below by $\left(L_{\rm Planck}\right)^{1/2}$ in supersymmetric theories and
by $\left(\frac{2}{3}L_{\rm Planck}\right)^{1/2}$ in non-supersymmetric theories, the infinite tower of light states
inevitably appears when $\phi$ travels much more than $(L_{\rm Planck})^{-1/2}$. On the other hand, since $\alpha_{\rm AdS}$ is bounded above by $(8 \pi L_{\rm AdS})^{1/2}$ (corresponding to $\alpha \leq 1$ in the CFT unit), $\phi$ must travel more than $(L_{\rm AdS})^{-1/2}$ in order to see this phenomenon.

Before we close this section, 
it may be worthwhile to point out that, although the proof of Theorems~\ref{thm:zerogaptoinfdist} and \ref{thm:gapvanishing} in section~\ref{sec:zerogaptoinfdist} assumes the existence of an exactly marginal operator for each tangent vector on the conformal manifold ${\cal M}$, we did not use this assumption in the proof of Theorem~\ref{thm:CFTinthelimit} in this section. We have
only assumed that there is an infinite sequence of CFTs with a suitably identified common set of light operators whose conformal weights vanish in the limit and the four-point functions of these light operators are well-defined in the limit (see \cite{Roggenkamp:2003qp} for a detailed discussion of this general limiting procedure). Thus,  Theorem~\ref{thm:CFTinthelimit} may be applicable to a larger class of families of CFTs. For example, let us 
consider the $k\rightarrow \infty$ limit of the $A_k$-type Virasoro minimal model CFTs, where the central charge approaches $c=1$ and $\Delta_{\rm gap}$ vanishes as 
$1 - c \propto 1/k^2$. 
In  \cite{Runkel:2001ng}, it was shown that 
the limiting CFT is described by a non-conventional non-compact boson. More recently, it was found in  \cite{Mazel:2024alu} that there is a pair of walls in the target space with a tachyon-dilaton profile and that the distance between these walls becomes infinite in the $k\rightarrow \infty$ limit. These results appear consistent with Theorem~\ref{thm:CFTinthelimit} since the limiting CFT in this case is described by the $\mR$ sigma-model away from the walls.

\section{Discussion}
\label{sec:discussion}

In this paper, we have shown that for $d=2$ unitary compact CFTs, any point on the conformal manifold $\cM$ where
 the conformal dimension of a nontrivial primary operator vanishes 
 is at infinite distance with respect to the Zamolodchikov metric on $\cM$. We have discussed 
 how this limit is approached and derived the universal bounds $c^{-1/2} \leq \alpha \leq 1$ on the parameter $\A$ in the exponential decay of conformal dimensions $\Delta= \exp(-\A t+O(1))$ with respect to the geodesic distance $t$. We have also deduced universal
 properties of the limiting CFT such as the emergence of a large target space. Moreover, we have extracted the implications of our CFT results for mass decay on the moduli space of quantum gravity. 
The immediate question is whether the converse statement can be proven or falsified: namely, what happens when we travel infinite geodesic distance on the conformal manifold $\cM$?

Though we have derived these results assuming that there is a continuous family of unitary CFTs at the same central charge and four-point functions of light operators are well-defined in the limit, some of them can be proven under weaker assumptions. In particular, Theorem \ref{thm:CFTinthelimit} is applicable to discrete families of CFTs with varying central charges. This unlocks the potential to study flat space limits of AdS gravities (see also Section~\ref{sec:gravityimplications})
and  verify the conjecture in \cite{Lust:2019zwm}. It would be interesting to find out whether the other theorems also hold for discrete families with a suitable measure.  

In higher dimensions, Conjecture~\ref{conj:CFTI},
which is analogous to our Theorem~\ref{thm:zerogaptoinfdist}, 
has been proven for supersymmetric theories in \cite{Perlmutter:2020buo} and more recently for general CFTs in \cite{Baume:2023msm}. However except for special classes of supersymmetric theories \cite{Perlmutter:2020buo}, no universal bounds on the exponential rate (analogous to $\A$ in two dimensions) are known. Here our universal upper bound on $\A$ follows from the bootstrap equation for the four-point function, and it is natural to ask whether this idea is useful in higher dimensions. 
As a more direct application of the bootstrap philosophy to gravity, it may also be interesting to apply the $S$-matrix bootstrap (see \cite{Kruczenski:2022lot} for a recent review) to find constraints on the effective theory of massless scalar fields coupled to gravity in asymptotically flat spacetime.
Furthermore, it remains an open question to prove or falsify the converse, as for the $d=2$ CFT/$d=3$ gravity.

There are many other intriguing features of conformal manifolds that appear to be universal and worth further investigating. For example, in $d=2$, the singularities on the conformal manifolds seem to be one of the following four types with distinct features: orbifold point (enhanced symmetries), conifold point (continuous spectrum with a non-zero gap above the vacuum),\footnote{One may be tempted to characterize the conifold singularity in terms of curvature divergence in the Zamolodchikov metric, which happens for the quintic SCFT as we have seen in Section~\ref{sec:quinticexample}. However this latter feature depends on the central charge of the CFT and in particular the conifold points of the K3 SCFT have finite curvature. 
Instead a more intrinsic way to identify this singularity in general CFT is as stated here.} branching point (accidental exactly marginal operators), and infinite distance limit (vanishing gap).\footnote{Among the four possibilities, the conifold points and the infinite distance limits share the property that the CFT becomes singular there ($i.e$ develops a continuous spectrum). However, in addition to the differences already noted, a crucial distinction is that the conifold points are expected to be \textit{strongly} coupled, whereas the infinite distance limits are \textit{weakly} coupled (containing a decoupled free sector) as we have seen in Section~\ref{sec:limitingCFT}. It would be interesting to make this statement more precisely for the conifold points. We thank Cumrun Vafa for discussions on this point.} It would be interesting to understand if they are all the possibilities for $d=2$ conformal manifold and what are the global constraints on their existence. One could ask the same questions for higher dimensional CFTs. In particular, the sum rules derived in the recent paper \cite{Balthazar:2022hzb} relating Zamolodchikov curvature to OPE data in the CFT could be useful to address these questions.

Another potentially interesting direction is to study topological constraints on the conformal manifolds. 
In  \cite{Ooguri:2006in}, the following conjecture was proposed
in addition to Conjectures \ref{conj:zero}, \ref{conj:one}, and \ref{conj:two}.
\setcounter{conj}{2}
\renewcommand{\theconj}{\arabic{conj}}
\begin{conj}
There is no non-trivial $1$-cycle with minimum length within a given homotopy class in ${\cal M}$. 
\label{conj:1cycle}
\end{conj}
This conjecture has been proven for gravitational theories in flat space with more than eight supercharges \cite{Cecotti:2015wqa}.
Since it was motivated by the absence of global symmetries in quantum gravity, which has been proven in AdS by the consistency of CFT
\cite{Harlow:2018jwu,Harlow:2018tng},
it may be possible to prove 
Conjecture~\ref{conj:1cycle} for general AdS gravities similarly. 

Finally, it would be interesting to further develop and generalize the bootstrap analysis in
\cite{Lin:2015wcg,Lin:2016gcl} for $d=2$ CFTs and in \cite{Chester:2021aun,Chester:2023ehi} for $d=4$ to probe more refined CFT data over the conformal manifold,\footnote{It is also interesting to pursue a similar analysis for $d=3$ CFTs which admit conformal manifolds (see for example \cite{Baggio:2017mas}). However it is not known whether conformal manifolds in $d=3$ can admit infinite distance limits. See \cite{
Bobev:2021yya,Bobev:2023bxs} for potential 
examples of non-compact conformal manifolds from holography. We thank Nikolay Bobev for correspondence on this last point.} such as general constraints on the perturbative expansion around the infinite distance limit (cusp point) in terms of the anomalous dimensions of protected operators.

\section*{Acknowledgments}
We thank  Ofer Aharony, Bruno Balthazar, Nathan Benjamin, José Calderón-Infante, Sergio Cecotti, Thomas Hartman, Zohar Komargodski, Ying-Hsuan Lin, Juan Maldacena, Sridip Pal, Julio Parra-Martinez, Eric Perlmutter, Massimo Porrati, Daniel Roggenkamp, Thomas Rudelius, Shu-Heng Shao, Eva Silverstein, David Simmons-Duffin,  Yan Soibelman,  Cumrun Vafa, Irene Valenzuela, Katrin Wendland, Edward Witten, and Xi Yin for discussions. The work of
HO is supported in part by the U.S. Department of Energy, Office of Science, Office of High Energy Physics, under Award Number DE-SC0011632, JSPS Grants-in-Aid for Scientific Research 23K03379, the Bershadsky Fellowship,
the Guggenheim Fellowship, and the Simons Investigator Award (MPS-SIP-00005259).  
The work of YW is
supported in part by the NSF grant PHY-2210420 and by the Simons Junior Faculty Fellows program.
HO thanks the hospitalities of the Center for Cosmology and Particle Physics at New York University, where this work was initiated, and of the Center for the Fundamental Laws of Nature at Harvard University, where this work was completed. 
His work was also performed in part at
the Kavli Institute for the Physics and Mathematics of the Universe at the University of Tokyo, which is supported by the World Premier International Research Center Initiative, MEXT, Japan,  and at the Kavli Institute for Theoretical Physics (KITP) at the University of California, Santa Barbara,
which is supported by NSF grant PHY-2309135.

\bibliographystyle{JHEP}
\bibliography{ref} 

\end{document}